\documentclass[
twocolumn, 
 amsmath,amssymb,
 aps,
prl,
superscriptaddress
]{revtex4-2}

\usepackage{graphicx}
\usepackage{dcolumn}
\usepackage{bm}
\usepackage{times} 
\usepackage[colorlinks=true, allcolors=blue]{hyperref}
\usepackage{braket}
\usepackage{bbding}
\usepackage{natbib}
\usepackage{gensymb}
\usepackage{balance}
\usepackage{xcolor}

\begin{document}

\preprint{APS/123-QED}

\title{Electro-optic effects in some sliding ferroelectrics}

\author{Xueqing Wan}
\email{These authors contributed equally to this work}
\affiliation{%
Ministry of Education Key Laboratory for Nonequilibrium Synthesis and Modulation of Condensed Matter, Shaanxi Province Key Laboratory of Advanced Functional Materials and Mesoscopic Physics, School of Physics, Xi'an Jiaotong University, Xi'an 710049, China 
}%

\author{Zhenlong Zhang}
\email{These authors contributed equally to this work}
\affiliation{%
Ministry of Education Key Laboratory for Nonequilibrium Synthesis and Modulation of Condensed Matter, Shaanxi Province Key Laboratory of Advanced Functional Materials and Mesoscopic Physics, School of Physics, Xi'an Jiaotong University, Xi'an 710049, China 
}%

\author{Charles Paillard}
\affiliation{%
Smart Ferroic Materials Center, Physics Department and Institute for Nanoscience and Engineering, University of Arkansas, Fayetteville, Arkansas 72701, USA
}%
\affiliation{%
Universit\'{e} Paris-Saclay, CentraleSup\'{e}lec, CNRS, Laboratoire SPMS, 91190, Gif-sur-Yvette, France
}%

\author{Jinyang Ni}
\email{jyni@xjtu.edu.cn}
\affiliation{%
 Ministry of Education Key Laboratory for Nonequilibrium Synthesis and Modulation of Condensed Matter, Shaanxi Province Key Laboratory of Advanced Functional Materials and Mesoscopic Physics, School of Physics, Xi'an Jiaotong University, Xi'an 710049, China 
}%

\author{Lei Zhang}
\email{zhangleio@xjtu.edu.cn}
\affiliation{%
 Ministry of Education Key Laboratory for Nonequilibrium Synthesis and Modulation of Condensed Matter, Shaanxi Province Key Laboratory of Advanced Functional Materials and Mesoscopic Physics, School of Physics, Xi'an Jiaotong University, Xi'an 710049, China 
}%

\author{Zhijun Jiang}
\email{zjjiang@xjtu.edu.cn}
\affiliation{%
 Ministry of Education Key Laboratory for Nonequilibrium Synthesis and Modulation of Condensed Matter, Shaanxi Province Key Laboratory of Advanced Functional Materials and Mesoscopic Physics, School of Physics, Xi'an Jiaotong University, Xi'an 710049, China }%
\affiliation{ State Key Laboratory of Surface Physics and Department of Physics, Fudan University, Shanghai 200433, China}

\author{Laurent Bellaiche}
\affiliation{%
Smart Ferroic Materials Center, Physics Department and Institute for Nanoscience and Engineering, University of Arkansas, Fayetteville, Arkansas 72701, USA
}%
\affiliation{%
Department of Materials Science and Engineering, Tel Aviv University, Ramat Aviv, Tel Aviv 6997801, Israel
}%




\begin{abstract}
Sliding ferroelectrics, which exhibit out-of-plane polarization arising from specific stacking rather than conventional ionic displacements, are new types of ferroelectrics whose underdeveloped physics needs to be explored. Here, we investigate the electro-optic (EO) response of these materials using first-principles calculations, focusing on ZrI$_{2}$ as a prototype. We reveal that, contrary to conventional ferroelectrics, the EO effect in ZrI$_{2}$ is dominated by its electronic contribution rather than the ionic one, which promises faster EO responses. Furthermore, both biaxial and uniaxial strains significantly enhance this response, and a universal-like linear relationship between the band gap and such response is discovered. We also report a large elasto-optic coefficient that is independent of biaxial strain. Similar large linear EO coefficients and properties are found in other sliding ferroelectrics, including different zirconium dihalides, as well as BN and BP bilayers. These findings highlight sliding ferroelectrics as highly promising candidates for ultrafast nonlinear optical devices and reveal unusual mechanisms.

\end{abstract}

\maketitle


The burgeoning field of interlayer sliding ferroelectricity, evidenced in van der Waals (vdW) bilayers like hexagonal boron nitride (\(\mbox{h-BN}\)) \cite{Lei2017,yasuda2021,vizner2021} and \(\mbox{WTe}_2\) \cite{Fei2018,yang2018}, presents a novel paradigm for ferroelectric materials. While conventional ferroelectricity (for example, in perovskite oxide like BaTiO\textsubscript{3} \cite{cohen1992}) arises from polar phonon mode displacements and associated hybridization leading to field-switchable polarization, the polarization in emerging sliding ferroelectrics is driven by the breaking of spatial inversion symmetry through artificial stacking engineering.

Notably, recent experimental works have confirmed out-of-plane polarization in layered \(\beta\)-ZrI$_{2}$ (\(Pmn2_1\) space group) \cite{Ma2021, Ding2021, Corbett1982}, which originates from the interlayer sliding analogous to sliding ferroelectrics such as \(\mbox{h-BN}\) \cite{Lei2017,yasuda2021,vizner2021} and \(\mbox{WTe}_2\) \cite{Fei2018,yang2018}. The bulk phases of monoclinic \(\alpha\)-ZrI$_{2}$ (\(P2_1/m\) space group) and orthorhombic \(\beta\)-ZrI$_{2}$ have also been synthesized in experiment \cite{Corbett1982}. Both phases can be derived from the high-symmetry paraelectric parent phase of \textit{s}-ZrI$_{2}$ (\(Pnma\) space group) \cite{Corbett1982}. Crucially, the ferroelectric properties of \(\beta\)-ZrI$_{2}$ can be effectively controlled by mechanical means, including hydrostatic pressure, shear, and uniaxial strains \cite{Ding2021}, opening avenues for strain-engineered functionalities.

An important and unexplored question is whether the significantly different nature of sliding ferroelectrics also results in singularly different physical and functional responses. Among other properties, the linear electro-optic (EO) effect (or Pockels effect), which refers to the change in a material's refractive index when an external electric field is applied \cite{DrDomenico1969, Wemple1969} is highly relevant for potential applications in quantum and classical communications and fundamental optical devices, including bistable switches \cite{Martnez-Lorente2017} and optical resonators \cite{Guarino2007}. It is therefore urgent to address this question, which could significantly advance the new field of slide optoelectronics devoted to light-matter interactions in sliding ferroelectrics.

The main striking result of this Letter is the discovery that the origin of the EO response in sliding ferroelectrics is \textit{electronic} in nature. This result is in stark contrast to the EO response of conventional ferroelectrics, which is driven by the optical phonon response~\cite{Veithen2004, Paillard2019, Jiang2019}. Using first-principles calculations with a particular emphasis on interlayer-sliding ferroelectric \(\beta\)-ZrI$_{2}$, we further demonstrate that both biaxial and uniaxial strains enhance the electronic EO coefficient. Finally, our analysis reveals a strong linear correlation with the electronic band gap in both types of strain, which opens the door to the design of materials with giant electronic EO responses. Finally, a large and strain-independent {\it elasto-optic} coefficient is also discovered in ZrI$_{2}$. Remarkably, similar features are further predicted to occur in other sliding ferroelectrics, such as the whole family of zirconium dihalides, and bilayers made of BN or BP.

First-principles calculations based on density functional theory (DFT) are performed (see Supplemental Materials (SM) \cite{Supplemental_Material}) in ZrI$_{2}$ bulk. The optimized structures are used to calculate the EO tensor based on the density functional perturbation theory (DFPT) \cite{Veithen2005}. This method was proven to be highly accurate for predicting EO coefficients, as demonstrated for  \(\mbox{LiNbO}_{3}\) \cite{Veithen2004}, \(\mbox{BaTiO}_{3}\) \cite{Jiang2020, Veithen2004}, and \(\mbox{PbTiO}_{3}\) \cite{Veithen2004}.

\begin{figure*}
    	\centering
    	\includegraphics[scale=0.8]{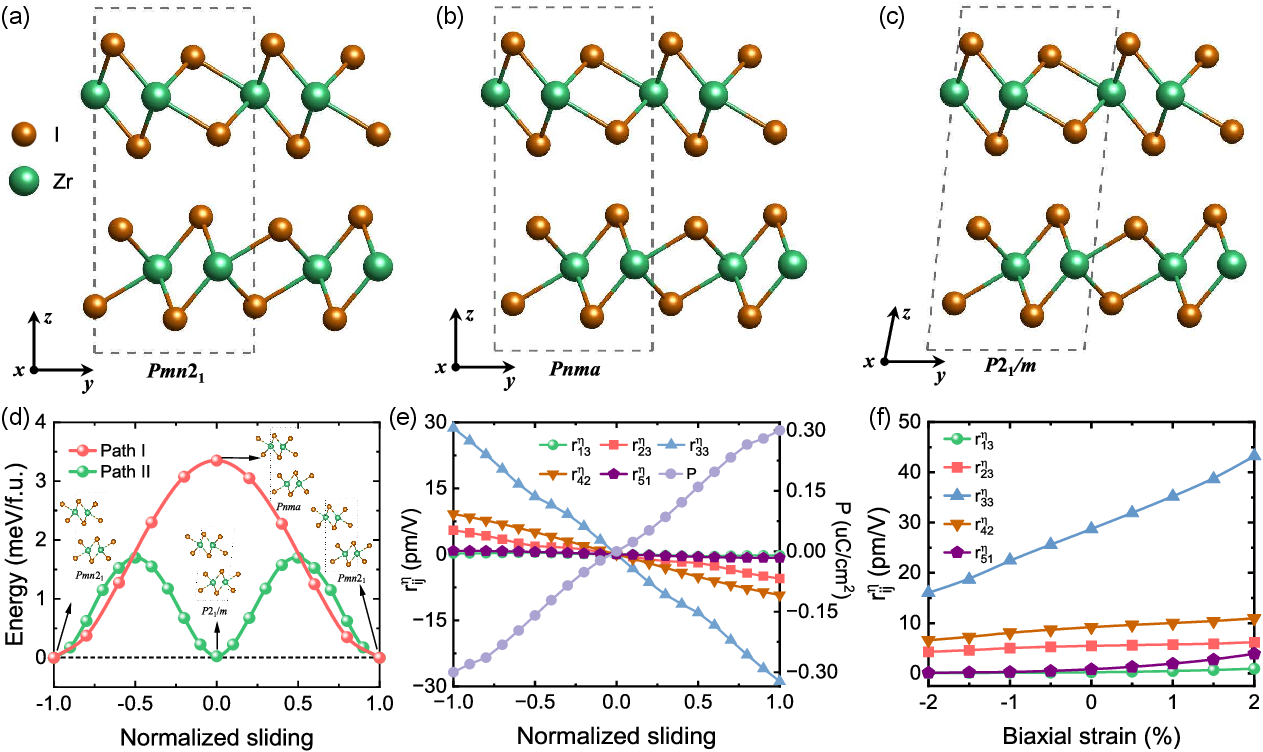}
    	\caption{The crystal structures of (a) \(\beta\)-ZrI\(_{2}\) (\(Pmn2_1\)), (b) \textit{s}-ZrI\(_{2}\) (\(Pnma\)), and (c) \(\alpha\)-ZrI\(_{2}\) (\(P2_1/m\)), respectively; (d) two possible polarization switching paths of ZrI\(_{2}\) bulk. The \textit{s}-ZrI\(_{2}\) and \(\alpha\)-ZrI\(_{2}\) are the intermediate states of path I and path II, respectively; (e) clamped EO tensor and polarization (\(P\)) as a function of the sliding path II; and (f) clamped EO tensor versus biaxial strain in \(\beta\)-ZrI$_{2}$.
    \label{fig1}}
\end{figure*}

In general, the linear (Pockels) EO tensor \(r_{ijk}\) can be expressed as \cite{Veithen2004, Veithen2005}:
\begin{equation} \label{Eq.1}
\Delta(\varepsilon^{-1})_{ij} = \sum_{k=1}^{3} r_{ijk} {\cal E}_{k},
\end{equation}

\noindent where \((\varepsilon^{-1})_{ij}\) refers to the inverse of the electronic dielectric tensor, while \({\cal E}_{k}\) represents the applied external electric field in the Cartesian direction \(k\).

The clamped EO tensor (zero strain that considers both electronic and ionic contributions while keeping the unit cell shape fixed) can be written as \cite{Veithen2004, Veithen2005}:
\begin{equation}\label{Eq.2}
r_{ijk}^{\mathrm{\eta}} = r_{ijk}^{\mathrm{\textrm{ele}}} + r_{ijk}^{\mathrm{\textrm{ion}}} = \frac{-8\pi}{n_{i}^{2} n_{j}^{2}} \chi_{ijk}^{(2)} - \frac{4\pi}{n_{i}^{2} n_{j}^{2} \sqrt{\Omega_{0}}} \sum_{m} \frac{\alpha_{ij}^{m} p_{k}^{m}}{\omega_{m}^{2}},
\end{equation}

\noindent where \(r_{ijk}^{\mathrm{\textrm{ele}}}\) represents the bare electronic contribution, \(r_{ijk}^{\mathrm{\textrm{ion}}}\) denotes the ionic contribution, \(n_{i}\) and \(n_{j}\) are the principal refractive indices, \(\chi_{ijk}^{(2)}\) is the second-order nonlinear optical dielectric susceptibility, \(\Omega_{0}\) is the unit cell volume, \(\alpha_{ij}^{m}\) refers to the Raman susceptibility of mode \(m\), \(p_{k}^{m}\) indicates the polarity, and \(\omega_{m}\) is the transverse optic phonon mode frequency.

The unclamped (zero stress that adds a contribution involving the piezoelectricity) EO tensor is given by:
\begin{equation}\label{Eq.3}
r_{ijk}^{\mathrm{\sigma}} = r_{ijk}^{\eta} + \sum_{\alpha,\beta} p_{ij\alpha\beta} d_{k\alpha\beta},
\end{equation}

\noindent where \(p_{ij\alpha\beta}\) is the elasto-optic coefficient and \(d_{k\alpha\beta}\) is the piezoelectric strain coefficient \cite{Veithen2004, Veithen2005,Jiang2019}. Equations~\eqref{Eq.2} and \eqref{Eq.3} can be used to calculate the linear clamped and unclamped EO tensor from first principles \cite{Veithen2004, Jiang2019, Paillard2019, Paoletta2021, jiang2024_2D, Zhang2024}.

In addition, the elasto-optic tensor \( p_{ij\alpha\beta} \) can be calculated via \cite{zgonik1994}
\begin{equation}\label{Eq.4}
\Delta\left(\frac{1}{n^{2}}\right)_{ij} = \Delta\left(\varepsilon^{-1}\right)_{ij} = \sum_{\alpha,\beta} p_{ij\alpha\beta} x_{\alpha\beta},
\end{equation}

\noindent where \(n\) represents the refractive index and \(x_{\alpha\beta}\) is the strain tensor.

As shown in Figs.~\ref{fig1}(a)-\ref{fig1}(c), three different phases are considered for ZrI$_{2}$ bulk. They adopt symmetries of \(Pmn2_1\), \(P2_1/m\), and \(Pnma\), and are denoted as \(\beta\)-ZrI$_{2}$ \cite{Corbett1982}, \(\alpha\)-ZrI$_{2}$ \cite{guthrie1981}, and \textit{s}-ZrI$_{2}$ \cite{Ding2021}, respectively. The \(\beta\)-ZrI$_{2}$ is the ferroelectric ground state of ZrI$_{2}$ bulk, while \(\alpha\)-ZrI$_{2}$ and \textit{s}-ZrI$_{2}$ are both paraelectric phases. Note that  \textit{s}-ZrI$_{2}$ is the common parent phase for both \(\alpha\)-ZrI$_{2}$ and \(\beta\)-ZrI$_{2}$, and that \(\alpha\)-ZrI$_{2}$ and \(\beta\)-ZrI$_{2}$ are sister phases. The optimized lattice parameters and band gap of \(\beta\)-ZrI$_{2}$ (\(a=3.75, b=6.86, c=14.80\) {\AA}, \(E_{0}=0.20 \) eV ), \textit{s}-ZrI$_{2}$ (\(a=3.75, b=6.85, c=14.89\) {\AA}, \(E_{0}=0.19 \) eV), \(\alpha\)-ZrI$_{2}$ (\(a=3.75, b =6.86, c=15.01\) {\AA}, \(E_{0}=0.18 \) eV) agree well with previous theoretical (\(\beta\)-ZrI$_{2}$ with \(a = 3.75, b = 6.87, c=14.80\) {\AA}, \(E_{0}=0.15 \) eV \cite{Ding2021}) and experimental (\(\beta\)-ZrI$_{2}$ with \(a = 3.74, b = 6.83, c=14.89\) {\AA} \cite{Corbett1982}) values (the comparisons of other phases are summarized in Table S1 of SM \cite{Supplemental_Material}).  The planar-averaged differential charge density of \(\beta\)-ZrI$_{2}$ indicates an interlayer charge transfer from one side to the opposite side, causing the occurrence of an out-of-plane polarization (see Fig. S1 of the SM \cite{Supplemental_Material}). The calculated spontaneous electric polarization in \(\beta\)-ZrI$_{2}$ is equal to \(0.3\ \mu\text{C/cm}^{2}\), which is in agreement with previous theoretical values of \(0.39\ \mu\text{C/cm}^{2}\) \cite{Ding2021} and \(0.24\ \mu\text{C/cm}^{2}\) \cite{Ma2021}. Note that 
 this spontaneous polarization is much larger than many other sliding ferroelectrics, such as bilayer WTe$_{2}$ (\(0.054\ \mu\text{C/cm}^{2}\)) \cite{Liu2019} and 1\textit{T}$^{\prime \prime}$-\(\mbox{MoSe}_2\) (\(0.018\ \mu\text{C/cm}^{2}\)) \cite{Tang2025}.

Previous studies have demonstrated that the \(\beta\)-ZrI$_{2}$ and \(\alpha\)-ZrI$_{2}$ phases are dynamically stable, while \textit{s}-ZrI$_{2}$ lacks dynamical stability at 0 K due to two imaginary frequencies located around the \(\Gamma\) point  \cite{Ma2021, Ding2021}. Moreover, two ferroelectric switching paths are calculated, as shown in Fig.~\ref{fig1}(d). These two paths are obtained by the interlayer sliding of vdW layers [see the insets of Fig.~\ref{fig1}(d)]  along the \textit{y}-axis with this sliding being normalized to one for the final state and with \text{\textit{s}}-ZrI$_{2}$ and \(\alpha\)-ZrI$_{2}$ being the intermediate phases in paths I and II, respectively. It is significant that the path I has a higher energy barrier (\(\sim\)3.45 meV/f.u.) than the path II. The path II leads to a double-energy-peaks barrier \(\sim\)1.7 meV/f.u. and a very small energy difference of \(\sim\)0.01 meV/f.u. between \(\alpha\)-ZrI$_{2}$ and \(\beta\)-ZrI$_{2}$. Since the path I has a much larger energy barrier and that \textit{s}-ZrI$_{2}$ is dynamically unstable, only path II is considered in the following.

\begin{figure}
    	\centering
    	\includegraphics[scale=0.3]{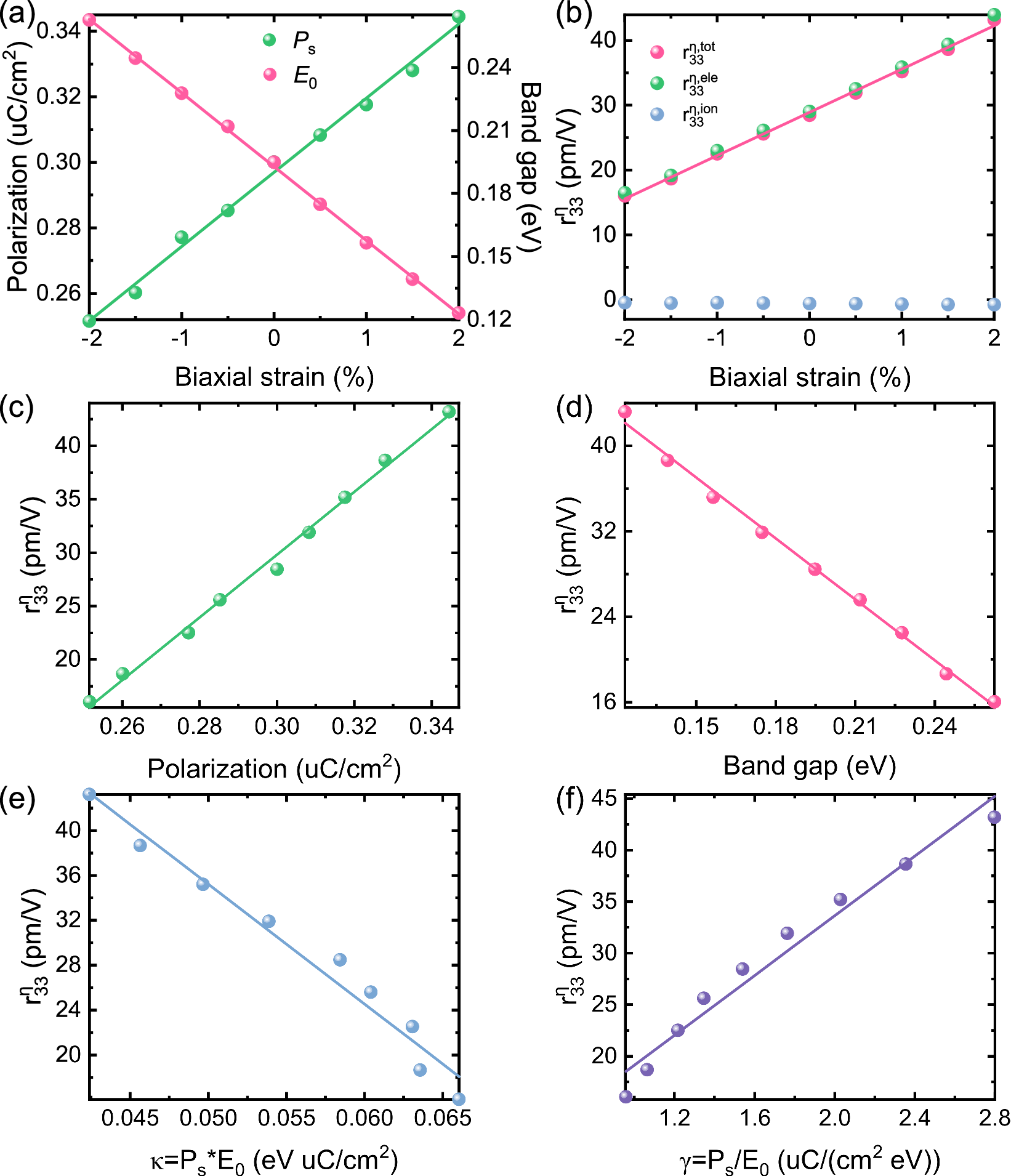}
    	\caption{(a) The polarization, band gap, and (b) clamped EO coefficient \(r_{33}^{\eta}\) as a function of biaxial strain. The clamped EO coefficient \(r_{33}^{\eta}\) as a function of (c) polarization, (d) band gap, (e) \(\kappa\) (the product between the polarization \(P_\text{s}\) and \(E_{0}\)), and (f) \(\gamma\) (the ratio between \(P_\text{s}\) and \(E_{0}\)) under biaxial strain.
    \label{fig2}}
\end{figure}

The clamped EO tensor and polarization as a function of normalized interlayer sliding (path II) are shown in Fig.~\ref{fig1}(e). The EO tensor of \(\beta\)-ZrI$_{2}$ bulk (${mm2}$ point group) has five independent elements in the Voigt notation \cite{Nye1985}: \(r_{13}^{\eta}\), \(r_{23}^{\eta}\), \(r_{33}^{\eta}\), \(r_{42}^{\eta}\), and \(r_{51}^{\eta}\). For \(\beta\)-ZrI$_{2}$ bulk, the calculated clamped EO coefficients are \(r_{13}^{\eta} = 0.2\) pm/V, \(r_{23}^{\eta} = 5.5\) pm/V, \(r_{33}^{\eta} = 28.7\) pm/V, \(r_{42}^{\eta} = 9.2\) pm/V, and \(r_{51}^{\eta} = 0.8\) pm/V. The clamped EO coefficient \(r_{33}^{\eta} = 28.7\) pm/V is particularly significant since it is very close to the one of the currently most used EO material, LiNbO$_{3}$ (that has a \(r_{33}^{\eta} = 30.8\) pm/V) \cite{Weber2002, Veithen2004}. Note that the EO tensor and polarization can be reversed via the interlayer sliding, as shown in Fig.~\ref{fig1}(e). In addition, several recent studies indicate that epitaxial strain can effectively regulate the value of polarization, band gap, and EO response \cite{Paillard2019, Jiang2019, Zhang2024}. This explains why Fig.~\ref{fig1}(f) further shows the presently computed effect of epitaxial biaxial strain on the EO response of bulk \(\beta\)-ZrI$_{2}$. Here, the biaxial strain ranges between ${-}$2\% and ${+}$2\%.  One can see in particular that \(r_{33}^{\eta}\) is very sensitive to these biaxial strains since it varies from 16.1 to 43.2 pm/V in this interval. This latter predicted EO coefficient \(r_{33}^{\eta}\) at 2\% strain is therefore larger than the experimental results in the \(R3c\) phase of \(\mbox{LiNbO}_{3}\) (\(r_{33}^{\eta} = 30.8\) pm/V), \(P4mm\) phase of \(\mbox{PbTiO}_{3}\) (\(r_{33}^{\eta} = 5.9\) pm/V) and \(\mbox{BaTiO}_{3}\) (\(r_{33}^{\eta} = 28\) pm/V) \cite{Weber2002, Veithen2004}. As indicated in Eq.~\eqref{Eq.2}, the total clamped EO tensor \(r_{33}^{\eta,\text{tot}}\) can be expressed as the sum of two contributions, namely electronic (\(r_{33}^{\eta,\text{ele}}\)) and ionic (\(r_{33}^{\eta,\text{ion}}\)) parts. It is a striking feature that the main contribution to \(r_{33}^{\eta,\text{tot}}\) in \(\beta\)-ZrI\(_{2}\) is of electronic origin \(r_{33}^{\eta,\text{ele}}\) while the ionic degrees of freedom \(r_{33}^{\eta,\text{ion}}\) contribute negligibly for all considered strains [see Fig.~\ref{fig2}(b)]. Therefore, the EO effect in these sliding ferroelectrics is of a fundamentally different nature than EO responses of conventional ion-displacement-type ferroelectrics, for which ionic contributions make the bulk of their response~\cite{Veithen2004, Jiang2020, Jiang2022, Zhang2024, Paillard2019, delodovici2023}. It can thus be hoped that ultrafast EO modulation may be achieved in these sliding ferroelectrics, albeit the small band gap of ZrI\textsubscript{2} may be a cause for undesired strong optical absorption. In that respect, ZrCl\textsubscript{2} or ZrBr\textsubscript{2} may be preferable owing to their twice as large band gap and thus reduced optical absorption at common telecommunication wavelengths, meanwhile exhibiting a similar, mostly electronic EO response (see the SM \cite{Supplemental_Material}). This should also stimulate the search for high-bandgap electro-optic sliding ferroelectrics. Results about BN and BP bilayers reported in the SM \cite{Supplemental_Material} are particularly promising in that regard.

\begin{figure}
    \centering
    \includegraphics[scale=0.5]{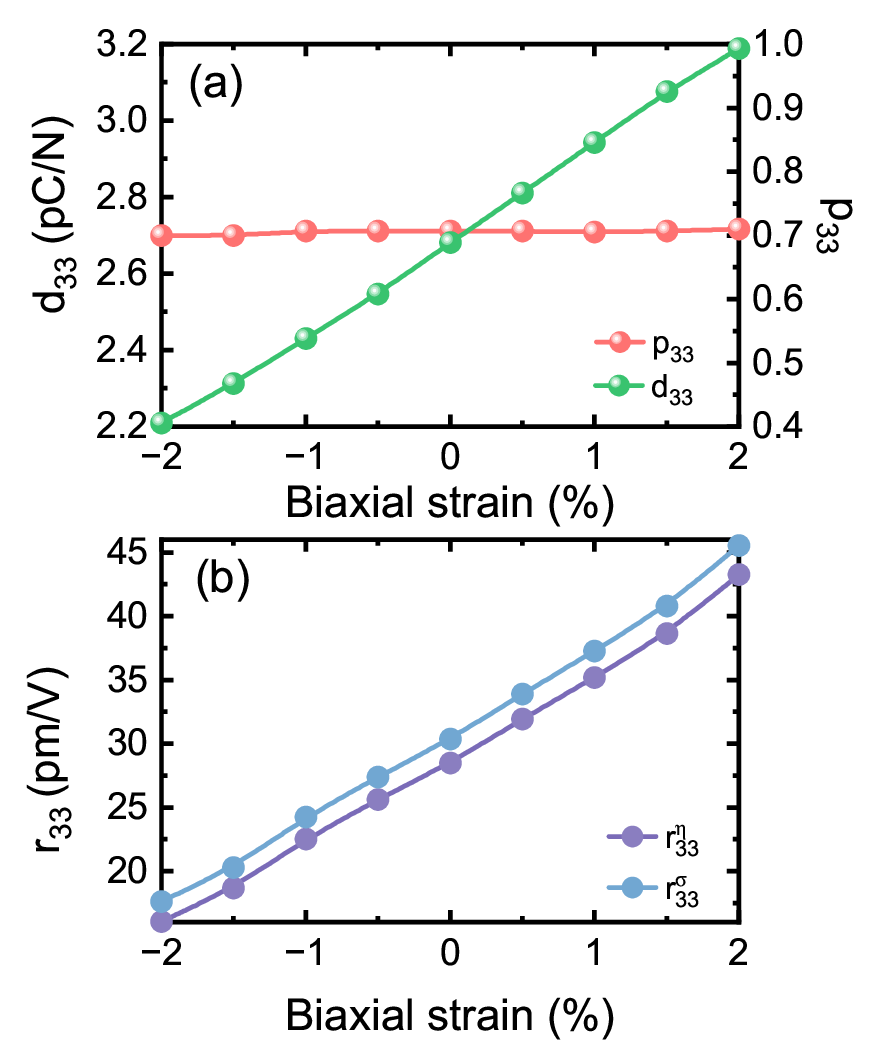}
    \caption{(a) Piezoelectric coefficient \(d_{33}\) and elasto-optic coefficient \(p_{33}\), and (b) clamped \(r_{33}^{\eta}\) and unclamped \(r_{33}^{\sigma}\) EO coefficients of \(\beta\)-ZrI\(_2\) as a function of biaxial strain.}
    \label{fig3}
\end{figure}

Since \(r_{33}^{\eta,\text{ele}}\) is proportional to the second-order nonlinear optical dielectric susceptibility \(\chi_{33}^{(2)}\) and the fourth power of refractive indice $-\frac{1}{{n_3}^4}$ and that this latter quantity is in first approximation independent on strain (see Figs.~S9-S10 of the SM~\cite{Supplemental_Material}), it is relevant to recall that Ref.~\cite{Ding2024} noticed a relationship between the second-harmonic generation and the product between the electronic band gap and the polarization (\(\kappa = E_{0}P_\text{s}\)) in two-dimensional ferroelectric NbOCl$_{2}$ monolayer.  A linear correlation is also indeed found here in our case for \(r_{33}^{\eta}\) as seen in Fig.~\ref{fig2}(e), namely the clamped EO coefficient \(r_{33}^{\eta}\) nearly linearly decreases with \(\kappa\). Practically, the data of \(r_{33}^{\eta}\) can be fitted by \(r_{33}^{\eta}\ = a + b\kappa\), where \(a = 88.6\) pm/V and \(b = -0.1 \, \text{cm}^3/\text{eV}^2 \cdot \text{C}
\), respectively. 
 In other words, a smaller  \(\kappa\) means a larger \(r_{33}^{\eta}\), resulting in \(r_{33}^{\eta}\) increasing with the biaxial strain since \(\kappa\) decreases as the biaxial strain increases from ${-}$2\% to ${+}$2\% [see Fig.~\ref{fig2}(e)]. Such latter decrease is due to the fact that for biaxial strains ranging between ${-}$2\% and ${+}$2\%, the electronic band gap (the difference in energy between the conduction band minimum and valence band maximum) strongly decreases by about 54\% from 0.26\ to 0.12\ eV [see Fig.~\ref{fig2}(a)], while the polarization ``only'' increases by about 36\% from 0.25\ to \(0.34\ \mu\text{C/cm}^{2}\) (within the phase associated with the ferroelectric \(Pmn2_1\) space group). Note that this polarization enhancement reflects the increasing difference in charge density between the two vdW layers forming each unit cell, when the biaxial strain varies from ${-}$2\% to ${+}$2\% (see Fig.~S1~\cite{Supplemental_Material}). We also note that the EO coefficient of \(\beta\)-ZrI$_{2}$ exhibits a linear relationship with the band gap because the second-order nonlinear optical susceptibility varies linearly with the band gap (see Fig.~S15~\cite{Supplemental_Material}).
     
  Let us also briefly mention the case of the {\it unclamped} EO coefficients (that adds a piezoelectric contribution) for \(\beta\)-ZrI\(_{2}\) under biaxial strains. As shown in Fig.~\ref{fig3}(b), both the clamped and unclamped EO coefficients exhibit a mostly linear behavior with values close to each other as a function of biaxial strain in \(\beta\)-ZrI\(_{2}\). This is because, as demonstrated in Fig.~\ref{fig3}(a), the piezoelectric strain coefficient \(d_{33}\) is really small, with values varying from 2.2\ to 3.2\ pC/N, even if the mostly strain-independent elasto-optic coefficient \(p_{33}\) is rather large. As a matter of fact,  its value is 0.7, which is about 12 times larger in magnitude than in trigonal LiNbO$_{3}$ for which \(p_{33}=0.06 \) \cite{Weber2002}. This property makes it highly promising for developing advanced technologies, such as tunable photonic devices \cite{saleh2019} and optomechanical sensors \cite{eichenfield2009}.

\begin{figure}

    	\centering
    	\includegraphics[scale=0.3]
        {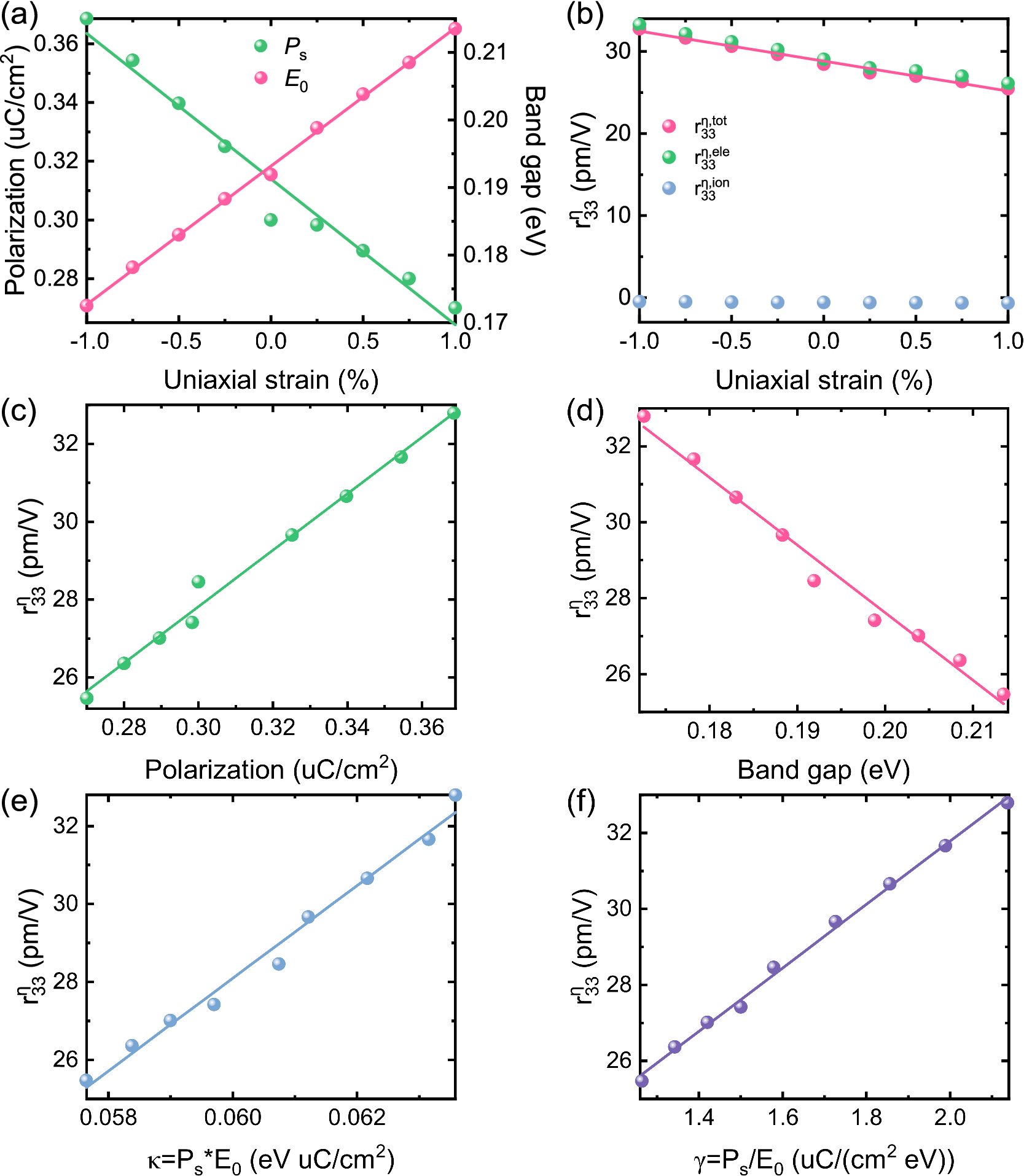}
    	\caption{Same as Fig.~\ref{fig2} but under uniaxial strain.
    \label{fig4}}
\end{figure}

Let us finally investigate \(\beta\)-ZrI\(_{2}\) under {\it uniaxial} strain (applied along the spontaneous polarization \textit{z}-axis) ranging between ${-}$1\% and ${+}$1\%. The polarization, band gap, and clamped EO coefficient \(r_{33}^{\eta}\) (including the total, electronic, and ionic contributions) relationships under uniaxial strain are shown in Figs.~\ref{fig4}(a)-\ref{fig4}(f). Interestingly, Fig.~\ref{fig4}(a) shows that uniaxial strain has the opposite qualitative effect on these properties than biaxial strain. As a matter of fact, the polarization now linearly decreases from 0.37\ to \(0.27\ \mu\text{C/cm}^{2}\) for uniaxial strains ranging between ${-}$1\% and ${+}$1\%, while the band gap linearly and concomitantly increases with value varying from 0.17\ to 0.21\ eV. Furthermore, the total clamped EO coefficient \(r_{33}^{\eta,\text{tot}}\) now linearly decreases with uniaxial strain from 32.8\ to 25.5\ pm/V---as indicated by Fig.~\ref{fig4}(b). As shown in Fig.~\ref{fig4}(e), the clamped EO coefficient \(r_{33}^{\eta}\) still adopts a linear relationship with \(\kappa\), but this relationship is now such that \(r_{33}^{\eta}\) increases linearly with \(\kappa\), in contrast with the case of biaxial strains that revealed a linear decrease. For the uniaxial case, this linear equation is given by $r_{33}^{\eta}\ = a + b\kappa$ with $a = -43.2$ pm/V and $b = 0.1\,\text{cm}^3/\text{eV}^2 ~ \text{C}$, respectively. In fact, such opposite behavior of  \(r_{33}^{\eta}\) as a function of  \(\kappa\) in the biaxial {\it versus} uniaxial strains mostly arises from the opposite behaviors of \(\chi_{33}^{(2)}\) with \(\kappa\) in these two different strain cases, as demonstrated in Figs.~S11 and S12 of the SM \cite{Supplemental_Material}. 

One must thus conclude that \(\chi_{33}^{(2)}\) and $r_{33}^{\eta}$ can indeed be linearly dependent on \(\kappa\), not only the magnitude but also sign of the slope of these straight lines depends on the chosen physical handle, e.g., biaxial strain, uniaxial strain, etc.
In fact, Ref.~\cite{Ding2024} itself found two different linear behaviors of  \(\chi_{33}^{(2)}\) as a function of $\kappa$ depending on the sign of their strain along the \textit{y}-axis: a linear increase for negative strains {\it versus} a linear decrease for positive strains, hence the name ``volcano'' relationship given by these authors. 

One may thus wonder if there are other quantities that correlate better with $r_{33}^{\eta}$ for these two different kinds of strains. Let us pay attention to three of these quantities, namely the \(P_\text{s}\) absolute value of the polarization, the $E_{0}$ band gap but also another combined parameter \(\gamma\) that we introduce here as the ratio between \(P_\text{s}\) and $E_{0}$. Figure~\ref{fig2}(c) shows  \(r_{33}^{\eta}\) as a function of  \(P_\text{s}\) for the {\it biaxial strain} case, which yields a linear increase with an intercept of \(-58.3\) \(\text{pm/V}\) and a slope of \(2.9 \times 10^{-8}\) \(\text{m}^{3}/\text{C}\,\cdot\text{V}\). An increase is also found for this function in the {\it uniaxial strain} situation but with a different intercept of \(6.08\) \(\text{pm/V}\) and a slope about four times smaller of \(7.2\times 10^{-9}\)  \(\text{m}^{3}/\text{C}\,\cdot\text{V}\) as shown in Fig.~\ref{fig4}(c). On the other hand, a similar quantitative correlation exists between $r_{33}^{\eta}$ and the band gap in both strain cases: Fig.~\ref{fig2}(d) reveals that \(r_{33}^{\eta}\) linearly decreases as the band gap increases with an intercept of \(65.5\) \(\text{pm/V}\) and a slope of \(-1.9 \times 10^{-10}\)~\(\text{m}\,\textit{e}^{-1}\,\text{V}^{-2}\) for biaxial strains, which is really close to the corresponding linear decrease displayed in Fig.~\ref{fig4}(d) for uniaxial strains yielding an intercept of \(62.9\) \(\text{pm/V}\) and a slope of \(-1.7 \times 10^{-10}\) ~ \(\text{m}\,\textit{e}^{-1}\,\text{V}^{-2}\). Regarding the behavior of $r_{33}^{\eta}$ as a function of $\gamma$,  Figs.~\ref{fig2}(f) and~\ref{fig4}(f) indicate that both cases of strains provide a linearly increasing function but with a different intercept  (4.6 \text{pm/V} {\it versus} 15.08 \text{pm/V}  for biaxial and uniaxial strains, respectively) and a different slope (\(1.4\times 10^{-15}~\text{m}^3 ~ \textit{e} / \mu \text{C}\) ~ {\it versus}~ \(0.8\times 10^{-15} ~\text{m}^3 ~ \textit{e} / \mu \text{C}\) for biaxial and uniaxial strains, respectively). It thus appears that the only ``universal'' qualitative and quantitative relation we found for both cases of the biaxial and uniaxial strains is that between $r_{33}^{\eta}$ and the band gap. Such universality-like behavior is also found for other zirconium dihalides, as shown in the SM~\cite{Supplemental_Material}.

 It is also important to realize that Fig.~\ref{fig4}(b) indicates that \(r_{33}^{\eta,\text{tot}}\) still comes from the electronic contribution (\(r_{33}^{\eta,\text{ele}}\)) while the ionic part (\(r_{33}^{\eta,\text{ion}}\)) is almost zero for all considered uniaxial strains. One can thus safely conclude that the EO effects in sliding ferroelectrics, both for clamped and unclamped cases, but also under stress-free, biaxial, and uniaxial strain situations, originate from the electronic contribution that is directly proportional to the second-order nonlinear optical dielectric susceptibility. In fact, the electronic nature of the EO response appears to be general for sliding systems, since we also numerically found it in BN and BP bilayers (see the SM \cite{Supplemental_Material}). These characteristics should enable ultrafast response speeds, such as in bilayer boron nitride with room-temperature operation of ferroelectric field effect transistor, which exhibits switching speeds on the nanosecond scale and high endurance exceeding \(10^{11}\) switching cycles~\cite{Yasuda2024}. Furthermore, the sliding ferroelectric of bilayer 3R-MoS\(_{2}\) also possesses a stronger substantial fatigue effect under the electric field~\cite{Bian2024}. These advantages over conventional ferroelectrics may lead to breakthroughs in using sliding ferroelectrics for high-performance electro-optic devices.

In this study, we performed detailed first-principles calculations to investigate the linear EO response and related effects in the sliding ferroelectric \(\beta\)-ZrI\(_{2}\). Large clamped EO and elasto-optic coefficients are predicted. Significantly, the EO effect in \(\beta\)-ZrI\(_{2}\) primarily originates from its electronic contribution, in stark contrast to conventional ferroelectrics. This unique mechanism promises faster electrical modulation of the optical index. We further show that both biaxial and uniaxial strains can effectively control the magnitude of the EO response. Remarkably, we demonstrate a universal linear dependence of the electro-optic coefficient on the band gap under both biaxial and uniaxial strains, a finding that will undoubtedly inform the discovery of materials with giant \textit{electronic} (and thus ultrafast compared to the slow dynamics of optical phonons) EO responses. The SM~\cite{Supplemental_Material} further confirms these remarkable features in \(\beta\)-ZrBr\(_{2}\) and \(\beta\)-ZrCl\(_{2}\), as well as in BN and BP bilayers, and suggests that this might be a universal feature of sliding ferroelectrics which will bolster further theoretical and experimental exploration.

This work is supported by the National Natural Science Foundation of China (Grants No.\,12374092 and No.\,T2425029), the Natural Science Basic Research Program of Shaanxi (Grants No.\,2023-JC-YB-017 and No.\,2022JC-DW5-02), the Shaanxi Fundamental Science Research Project for Mathematics and Physics (Grant No.\,22JSQ013), the Young Talent Support Plan of Xi'an Jiaotong University, the Open Project of the State Key Laboratory of Surface Physics (Grant No.\,KF2023\_06), and the Xiaomi Young Talents Program. C.P. and L.B. thank  the Award No. FA9550-23-1-0500 from the U.S. Department of Defense under the DEPSCoR program. L.B. also acknowledges the MonArk NSF Quantum Foundry supported by the National Science Foundation Q-AMASE-i Program under NSF Award No. DMR-1906383, the ARO Grant No. W911NF-21--1--0113, and the Vannevar Bush Faculty Fellowship (VBFF) Grant No. N00014-20-1--2834 from the Department of Defense. The HPC Platform of Xi'an Jiaotong University is also acknowledged. 

\appendix

\nocite{*}
\bibliography{Main_text}

\begin{thebibliography}{53}%
\makeatletter
\providecommand \@ifxundefined [1]{%
 \@ifx{#1\undefined}
}%
\providecommand \@ifnum [1]{%
 \ifnum #1\expandafter \@firstoftwo
 \else \expandafter \@secondoftwo
 \fi
}%
\providecommand \@ifx [1]{%
 \ifx #1\expandafter \@firstoftwo
 \else \expandafter \@secondoftwo
 \fi
}%
\providecommand \natexlab [1]{#1}%
\providecommand \enquote  [1]{``#1''}%
\providecommand \bibnamefont  [1]{#1}%
\providecommand \bibfnamefont [1]{#1}%
\providecommand \citenamefont [1]{#1}%
\providecommand \href@noop [0]{\@secondoftwo}%
\providecommand \href [0]{\begingroup \@sanitize@url \@href}%
\providecommand \@href[1]{\@@startlink{#1}\@@href}%
\providecommand \@@href[1]{\endgroup#1\@@endlink}%
\providecommand \@sanitize@url [0]{\catcode `\\12\catcode `\$12\catcode `\&12\catcode `\#12\catcode `\^12\catcode `\_12\catcode `\%12\relax}%
\providecommand \@@startlink[1]{}%
\providecommand \@@endlink[0]{}%
\providecommand \url  [0]{\begingroup\@sanitize@url \@url }%
\providecommand \@url [1]{\endgroup\@href {#1}{\urlprefix }}%
\providecommand \urlprefix  [0]{URL }%
\providecommand \Eprint [0]{\href }%
\providecommand \doibase [0]{https://doi.org/}%
\providecommand \selectlanguage [0]{\@gobble}%
\providecommand \bibinfo  [0]{\@secondoftwo}%
\providecommand \bibfield  [0]{\@secondoftwo}%
\providecommand \translation [1]{[#1]}%
\providecommand \BibitemOpen [0]{}%
\providecommand \bibitemStop [0]{}%
\providecommand \bibitemNoStop [0]{.\EOS\space}%
\providecommand \EOS [0]{\spacefactor3000\relax}%
\providecommand \BibitemShut  [1]{\csname bibitem#1\endcsname}%
\let\auto@bib@innerbib\@empty
\bibitem [{\citenamefont {Li}\ and\ \citenamefont {Wu}(2017)}]{Lei2017}%
  \BibitemOpen
  \bibfield  {author} {\bibinfo {author} {\bibfnamefont {L.}~\bibnamefont {Li}}\ and\ \bibinfo {author} {\bibfnamefont {M.}~\bibnamefont {Wu}},\ }\bibfield  {title} {\bibinfo {title} {Binary compound bilayer and multilayer with vertical polarizations: two-dimensional ferroelectrics, multiferroics, and nanogenerators},\ }\href {https://doi.org/10.1021/acsnano.7b02756} {\bibfield  {journal} {\bibinfo  {journal} {ACS Nano}\ }\textbf {\bibinfo {volume} {11}},\ \bibinfo {pages} {6382} (\bibinfo {year} {2017})}\BibitemShut {NoStop}%
\bibitem [{\citenamefont {Yasuda}\ \emph {et~al.}(2021)\citenamefont {Yasuda}, \citenamefont {Wang}, \citenamefont {Watanabe}, \citenamefont {Taniguchi},\ and\ \citenamefont {Jarillo-Herrero}}]{yasuda2021}%
  \BibitemOpen
  \bibfield  {author} {\bibinfo {author} {\bibfnamefont {K.}~\bibnamefont {Yasuda}}, \bibinfo {author} {\bibfnamefont {X.}~\bibnamefont {Wang}}, \bibinfo {author} {\bibfnamefont {K.}~\bibnamefont {Watanabe}}, \bibinfo {author} {\bibfnamefont {T.}~\bibnamefont {Taniguchi}},\ and\ \bibinfo {author} {\bibfnamefont {P.}~\bibnamefont {Jarillo-Herrero}},\ }\bibfield  {title} {\bibinfo {title} {Stacking-engineered ferroelectricity in bilayer boron nitride},\ }\href {https://doi.org/10.1126/science.abd3230} {\bibfield  {journal} {\bibinfo  {journal} {Science}\ }\textbf {\bibinfo {volume} {372}},\ \bibinfo {pages} {1458} (\bibinfo {year} {2021})}\BibitemShut {NoStop}%
\bibitem [{\citenamefont {Vizner~Stern}\ \emph {et~al.}(2021)\citenamefont {Vizner~Stern}, \citenamefont {Waschitz}, \citenamefont {Cao}, \citenamefont {Nevo}, \citenamefont {Watanabe}, \citenamefont {Taniguchi}, \citenamefont {Sela}, \citenamefont {Urbakh}, \citenamefont {Hod},\ and\ \citenamefont {Ben~Shalom}}]{vizner2021}%
  \BibitemOpen
  \bibfield  {author} {\bibinfo {author} {\bibfnamefont {M.}~\bibnamefont {Vizner~Stern}}, \bibinfo {author} {\bibfnamefont {Y.}~\bibnamefont {Waschitz}}, \bibinfo {author} {\bibfnamefont {W.}~\bibnamefont {Cao}}, \bibinfo {author} {\bibfnamefont {I.}~\bibnamefont {Nevo}}, \bibinfo {author} {\bibfnamefont {K.}~\bibnamefont {Watanabe}}, \bibinfo {author} {\bibfnamefont {T.}~\bibnamefont {Taniguchi}}, \bibinfo {author} {\bibfnamefont {E.}~\bibnamefont {Sela}}, \bibinfo {author} {\bibfnamefont {M.}~\bibnamefont {Urbakh}}, \bibinfo {author} {\bibfnamefont {O.}~\bibnamefont {Hod}},\ and\ \bibinfo {author} {\bibfnamefont {M.}~\bibnamefont {Ben~Shalom}},\ }\bibfield  {title} {\bibinfo {title} {Interfacial ferroelectricity by van der waals sliding},\ }\href {https://doi.org/10.1126/science.abe8177} {\bibfield  {journal} {\bibinfo  {journal} {Science}\ }\textbf {\bibinfo {volume} {372}},\ \bibinfo {pages} {1462} (\bibinfo {year} {2021})}\BibitemShut {NoStop}%
\bibitem [{\citenamefont {Fei}\ \emph {et~al.}(2018)\citenamefont {Fei}, \citenamefont {Zhao}, \citenamefont {Palomaki}, \citenamefont {Sun}, \citenamefont {Miller}, \citenamefont {Zhao}, \citenamefont {Yan}, \citenamefont {Xu},\ and\ \citenamefont {Cobden}}]{Fei2018}%
  \BibitemOpen
  \bibfield  {author} {\bibinfo {author} {\bibfnamefont {Z.}~\bibnamefont {Fei}}, \bibinfo {author} {\bibfnamefont {W.}~\bibnamefont {Zhao}}, \bibinfo {author} {\bibfnamefont {T.~A.}\ \bibnamefont {Palomaki}}, \bibinfo {author} {\bibfnamefont {B.}~\bibnamefont {Sun}}, \bibinfo {author} {\bibfnamefont {M.~K.}\ \bibnamefont {Miller}}, \bibinfo {author} {\bibfnamefont {Z.}~\bibnamefont {Zhao}}, \bibinfo {author} {\bibfnamefont {J.}~\bibnamefont {Yan}}, \bibinfo {author} {\bibfnamefont {X.}~\bibnamefont {Xu}},\ and\ \bibinfo {author} {\bibfnamefont {D.~H.}\ \bibnamefont {Cobden}},\ }\bibfield  {title} {\bibinfo {title} {Ferroelectric switching of a two-dimensional metal},\ }\href {https://doi.org/10.1038/s41586-018-0336-3} {\bibfield  {journal} {\bibinfo  {journal} {Nature (London)}\ }\textbf {\bibinfo {volume} {560}},\ \bibinfo {pages} {336} (\bibinfo {year} {2018})}\BibitemShut {NoStop}%
\bibitem [{\citenamefont {Yang}\ \emph {et~al.}(2018)\citenamefont {Yang}, \citenamefont {Wu},\ and\ \citenamefont {Li}}]{yang2018}%
  \BibitemOpen
  \bibfield  {author} {\bibinfo {author} {\bibfnamefont {Q.}~\bibnamefont {Yang}}, \bibinfo {author} {\bibfnamefont {M.}~\bibnamefont {Wu}},\ and\ \bibinfo {author} {\bibfnamefont {J.}~\bibnamefont {Li}},\ }\bibfield  {title} {\bibinfo {title} {Origin of two-dimensional vertical ferroelectricity in $\mbox{WTe}_{2}$ bilayer and multilayer},\ }\href {https://doi.org/10.1021/acs.jpclett.8b03654} {\bibfield  {journal} {\bibinfo  {journal} {J. Phys. Chem. Lett.}\ }\textbf {\bibinfo {volume} {9}},\ \bibinfo {pages} {7160} (\bibinfo {year} {2018})}\BibitemShut {NoStop}%
\bibitem [{\citenamefont {Cohen}(1992)}]{cohen1992}%
  \BibitemOpen
  \bibfield  {author} {\bibinfo {author} {\bibfnamefont {R.~E.}\ \bibnamefont {Cohen}},\ }\bibfield  {title} {\bibinfo {title} {Origin of ferroelectricity in perovskite oxides},\ }\href {https://doi.org/10.1038/358136a0} {\bibfield  {journal} {\bibinfo  {journal} {Nature (London)}\ }\textbf {\bibinfo {volume} {358}},\ \bibinfo {pages} {136} (\bibinfo {year} {1992})}\BibitemShut {NoStop}%
\bibitem [{\citenamefont {Ma}\ \emph {et~al.}(2021)\citenamefont {Ma}, \citenamefont {Liu}, \citenamefont {Ren},\ and\ \citenamefont {Nikolaev}}]{Ma2021}%
  \BibitemOpen
  \bibfield  {author} {\bibinfo {author} {\bibfnamefont {X.}~\bibnamefont {Ma}}, \bibinfo {author} {\bibfnamefont {C.}~\bibnamefont {Liu}}, \bibinfo {author} {\bibfnamefont {W.}~\bibnamefont {Ren}},\ and\ \bibinfo {author} {\bibfnamefont {S.~A.}\ \bibnamefont {Nikolaev}},\ }\bibfield  {title} {\bibinfo {title} {Tunable vertical ferroelectricity and domain walls by interlayer sliding in $\beta$-$\mbox{ZrI}_{2}$},\ }\href {https://doi.org/10.1038/s41524-021-00648-9} {\bibfield  {journal} {\bibinfo  {journal} {npj Comput. Mater.}\ }\textbf {\bibinfo {volume} {7}},\ \bibinfo {pages} {177} (\bibinfo {year} {2021})}\BibitemShut {NoStop}%
\bibitem [{\citenamefont {Ding}\ \emph {et~al.}(2021)\citenamefont {Ding}, \citenamefont {Chen}, \citenamefont {Gui}, \citenamefont {You}, \citenamefont {Yao},\ and\ \citenamefont {Dong}}]{Ding2021}%
  \BibitemOpen
  \bibfield  {author} {\bibinfo {author} {\bibfnamefont {N.}~\bibnamefont {Ding}}, \bibinfo {author} {\bibfnamefont {J.}~\bibnamefont {Chen}}, \bibinfo {author} {\bibfnamefont {C.}~\bibnamefont {Gui}}, \bibinfo {author} {\bibfnamefont {H.}~\bibnamefont {You}}, \bibinfo {author} {\bibfnamefont {X.}~\bibnamefont {Yao}},\ and\ \bibinfo {author} {\bibfnamefont {S.}~\bibnamefont {Dong}},\ }\bibfield  {title} {\bibinfo {title} {Phase competition and negative piezoelectricity in interlayer-sliding ferroelectric $\mbox{ZrI}_{2}$},\ }\href {https://doi.org/10.1103/PhysRevMaterials.5.084405} {\bibfield  {journal} {\bibinfo  {journal} {Phys. Rev. Mater.}\ }\textbf {\bibinfo {volume} {5}},\ \bibinfo {pages} {084405} (\bibinfo {year} {2021})}\BibitemShut {NoStop}%
\bibitem [{\citenamefont {Corbett}\ and\ \citenamefont {Guthrie}(1982)}]{Corbett1982}%
  \BibitemOpen
  \bibfield  {author} {\bibinfo {author} {\bibfnamefont {J.~D.}\ \bibnamefont {Corbett}}\ and\ \bibinfo {author} {\bibfnamefont {D.~H.}\ \bibnamefont {Guthrie}},\ }\bibfield  {title} {\bibinfo {title} {A second infinite-chain form of zirconium diiodide ($\beta$) and its coherent intergrowth with $\alpha$-zirconium diiodide},\ }\href {https://pubs.acs.org/doi/pdf/10.1021/ic00135a009} {\bibfield  {journal} {\bibinfo  {journal} {Inorg. Chem.}\ }\textbf {\bibinfo {volume} {21}},\ \bibinfo {pages} {1747} (\bibinfo {year} {1982})}\BibitemShut {NoStop}%
\bibitem [{\citenamefont {DrDomenico~Jr}\ and\ \citenamefont {Wemple}(1969)}]{DrDomenico1969}%
  \BibitemOpen
  \bibfield  {author} {\bibinfo {author} {\bibfnamefont {M.}~\bibnamefont {DrDomenico~Jr}}\ and\ \bibinfo {author} {\bibfnamefont {S.~H.}\ \bibnamefont {Wemple}},\ }\bibfield  {title} {\bibinfo {title} {Oxygen-octahedra ferroelectrics. $\mathrm{I}$. theory of electro-optical and nonlinear optical effects},\ }\href {https://doi.org/10.1063/1.1657458} {\bibfield  {journal} {\bibinfo  {journal} {J. Appl. Phys.}\ }\textbf {\bibinfo {volume} {40}},\ \bibinfo {pages} {720} (\bibinfo {year} {1969})}\BibitemShut {NoStop}%
\bibitem [{\citenamefont {Wemple}\ and\ \citenamefont {DiDomenico~Jr}(1969)}]{Wemple1969}%
  \BibitemOpen
  \bibfield  {author} {\bibinfo {author} {\bibfnamefont {S.~H.}\ \bibnamefont {Wemple}}\ and\ \bibinfo {author} {\bibfnamefont {M.}~\bibnamefont {DiDomenico~Jr}},\ }\bibfield  {title} {\bibinfo {title} {Oxygen-octahedra ferroelectrics. $\mathrm{II}$. electro-optical and nonlinear-optical device applications},\ }\href {https://doi.org/10.1063/1.1657459} {\bibfield  {journal} {\bibinfo  {journal} {J. Appl. Phys.}\ }\textbf {\bibinfo {volume} {40}},\ \bibinfo {pages} {735} (\bibinfo {year} {1969})}\BibitemShut {NoStop}%
\bibitem [{\citenamefont {Mart{\'\i}nez-Lorente}\ \emph {et~al.}(2017)\citenamefont {Mart{\'\i}nez-Lorente}, \citenamefont {Parravicini}, \citenamefont {Brambilla}, \citenamefont {Columbo}, \citenamefont {Prati}, \citenamefont {Rizza}, \citenamefont {Agranat},\ and\ \citenamefont {DelRe}}]{Martnez-Lorente2017}%
  \BibitemOpen
  \bibfield  {author} {\bibinfo {author} {\bibfnamefont {R.}~\bibnamefont {Mart{\'\i}nez-Lorente}}, \bibinfo {author} {\bibfnamefont {J.}~\bibnamefont {Parravicini}}, \bibinfo {author} {\bibfnamefont {M.}~\bibnamefont {Brambilla}}, \bibinfo {author} {\bibfnamefont {L.}~\bibnamefont {Columbo}}, \bibinfo {author} {\bibfnamefont {F.}~\bibnamefont {Prati}}, \bibinfo {author} {\bibfnamefont {C.}~\bibnamefont {Rizza}}, \bibinfo {author} {\bibfnamefont {A.~J.}\ \bibnamefont {Agranat}},\ and\ \bibinfo {author} {\bibfnamefont {E.}~\bibnamefont {DelRe}},\ }\bibfield  {title} {\bibinfo {title} {Scalable electro-optic control of localized bistable switching in broad-area $\mbox{VCSELs}$ using reconfigurable funnel waveguides},\ }\href {https://doi.org/10.1103/PhysRevApplied.7.064004} {\bibfield  {journal} {\bibinfo  {journal} {Phys. Rev. Appl.}\ }\textbf {\bibinfo {volume} {7}},\ \bibinfo {pages} {064004} (\bibinfo {year} {2017})}\BibitemShut {NoStop}%
\bibitem [{\citenamefont {Guarino}\ \emph {et~al.}(2007)\citenamefont {Guarino}, \citenamefont {Poberaj}, \citenamefont {Rezzonico}, \citenamefont {Degl'Innocenti},\ and\ \citenamefont {G{\"u}nter}}]{Guarino2007}%
  \BibitemOpen
  \bibfield  {author} {\bibinfo {author} {\bibfnamefont {A.}~\bibnamefont {Guarino}}, \bibinfo {author} {\bibfnamefont {G.}~\bibnamefont {Poberaj}}, \bibinfo {author} {\bibfnamefont {D.}~\bibnamefont {Rezzonico}}, \bibinfo {author} {\bibfnamefont {R.}~\bibnamefont {Degl'Innocenti}},\ and\ \bibinfo {author} {\bibfnamefont {P.}~\bibnamefont {G{\"u}nter}},\ }\bibfield  {title} {\bibinfo {title} {Electro-optically tunable microring resonators in lithium niobate},\ }\href {https://doi.org/10.1038/nphoton.2007.93} {\bibfield  {journal} {\bibinfo  {journal} {Nat. Photon.}\ }\textbf {\bibinfo {volume} {1}},\ \bibinfo {pages} {407} (\bibinfo {year} {2007})}\BibitemShut {NoStop}%
\bibitem [{\citenamefont {Veithen}\ \emph {et~al.}(2004)\citenamefont {Veithen}, \citenamefont {Gonze},\ and\ \citenamefont {Ghosez}}]{Veithen2004}%
  \BibitemOpen
  \bibfield  {author} {\bibinfo {author} {\bibfnamefont {M.}~\bibnamefont {Veithen}}, \bibinfo {author} {\bibfnamefont {X.}~\bibnamefont {Gonze}},\ and\ \bibinfo {author} {\bibfnamefont {P.}~\bibnamefont {Ghosez}},\ }\bibfield  {title} {\bibinfo {title} {First-principles study of the electro-optic effect in ferroelectric oxides},\ }\href {https://doi.org/10.1103/PhysRevLett.93.187401} {\bibfield  {journal} {\bibinfo  {journal} {Phys. Rev. Lett.}\ }\textbf {\bibinfo {volume} {93}},\ \bibinfo {pages} {187401} (\bibinfo {year} {2004})}\BibitemShut {NoStop}%
\bibitem [{\citenamefont {Paillard}\ \emph {et~al.}(2019)\citenamefont {Paillard}, \citenamefont {Prokhorenko},\ and\ \citenamefont {Bellaiche}}]{Paillard2019}%
  \BibitemOpen
  \bibfield  {author} {\bibinfo {author} {\bibfnamefont {C.}~\bibnamefont {Paillard}}, \bibinfo {author} {\bibfnamefont {S.}~\bibnamefont {Prokhorenko}},\ and\ \bibinfo {author} {\bibfnamefont {L.}~\bibnamefont {Bellaiche}},\ }\bibfield  {title} {\bibinfo {title} {Strain engineering of electro-optic constants in ferroelectric materials},\ }\href {https://doi.org/10.1038/s41524-018-0141-4} {\bibfield  {journal} {\bibinfo  {journal} {npj Comput. Mater.}\ }\textbf {\bibinfo {volume} {5}},\ \bibinfo {pages} {6} (\bibinfo {year} {2019})}\BibitemShut {NoStop}%
\bibitem [{\citenamefont {Jiang}\ \emph {et~al.}(2019)\citenamefont {Jiang}, \citenamefont {Paillard}, \citenamefont {Vanderbilt}, \citenamefont {Xiang},\ and\ \citenamefont {Bellaiche}}]{Jiang2019}%
  \BibitemOpen
  \bibfield  {author} {\bibinfo {author} {\bibfnamefont {Z.}~\bibnamefont {Jiang}}, \bibinfo {author} {\bibfnamefont {C.}~\bibnamefont {Paillard}}, \bibinfo {author} {\bibfnamefont {D.}~\bibnamefont {Vanderbilt}}, \bibinfo {author} {\bibfnamefont {H.}~\bibnamefont {Xiang}},\ and\ \bibinfo {author} {\bibfnamefont {L.}~\bibnamefont {Bellaiche}},\ }\bibfield  {title} {\bibinfo {title} {Designing multifunctionality via assembling dissimilar materials: Epitaxial $\mbox{AlN/ScN}$ superlattices},\ }\href {https://doi.org/10.1103/PhysRevLett.123.096801} {\bibfield  {journal} {\bibinfo  {journal} {Phys. Rev. Lett.}\ }\textbf {\bibinfo {volume} {123}},\ \bibinfo {pages} {096801} (\bibinfo {year} {2019})}\BibitemShut {NoStop}%
\bibitem [{Sup()}]{Supplemental_Material}%
  \BibitemOpen
  \href@noop {} {\bibinfo  {journal} {See Supplemental Material at ... for more details about (i) the computation details, planar-average charge density difference, and dynamics stability of \(\beta\)-ZrI\(_{2}\), \(\beta\)-ZrBr\(_{2}\), and \(\beta\)-ZrCl\(_{2}\), (ii) electro-optic (EO) properties of \(\beta\)-ZrI\(_{2}\), \(\beta\)-ZrBr\(_{2}\), and \(\beta\)-ZrCl\(_{2}\) under biaxial and uniaxial strains, (iii) piezoelectric coefficient, elasto-optic, and unclamped EO coefficient in \(\beta\)-ZrBr\(_{2}\) and \(\beta\)-ZrCl\(_{2}\), and (iv) EO effects in BN and BP bilayers, which includes Refs. [8,9,18, 21, 24, 27, 36-53]}\ }\BibitemShut {NoStop}%
\bibitem [{\citenamefont {Veithen}\ \emph {et~al.}(2005)\citenamefont {Veithen}, \citenamefont {Gonze},\ and\ \citenamefont {Ghosez}}]{Veithen2005}%
  \BibitemOpen
\bibfield  {journal} {  }\bibfield  {author} {\bibinfo {author} {\bibfnamefont {M.}~\bibnamefont {Veithen}}, \bibinfo {author} {\bibfnamefont {X.}~\bibnamefont {Gonze}},\ and\ \bibinfo {author} {\bibfnamefont {P.}~\bibnamefont {Ghosez}},\ }\bibfield  {title} {\bibinfo {title} {Nonlinear optical susceptibilities, raman efficiencies, and electro-optic tensors from first-principles density functional perturbation theory},\ }\href {https://doi.org/10.1103/PhysRevB.71.125107} {\bibfield  {journal} {\bibinfo  {journal} {Phys. Rev. B}\ }\textbf {\bibinfo {volume} {71}},\ \bibinfo {pages} {125107} (\bibinfo {year} {2005})}\BibitemShut {NoStop}%
\bibitem [{\citenamefont {Jiang}\ \emph {et~al.}(2020)\citenamefont {Jiang}, \citenamefont {Paillard}, \citenamefont {Xiang},\ and\ \citenamefont {Bellaiche}}]{Jiang2020}%
  \BibitemOpen
  \bibfield  {author} {\bibinfo {author} {\bibfnamefont {Z.}~\bibnamefont {Jiang}}, \bibinfo {author} {\bibfnamefont {C.}~\bibnamefont {Paillard}}, \bibinfo {author} {\bibfnamefont {H.}~\bibnamefont {Xiang}},\ and\ \bibinfo {author} {\bibfnamefont {L.}~\bibnamefont {Bellaiche}},\ }\bibfield  {title} {\bibinfo {title} {Linear versus nonlinear electro-optic effects in materials},\ }\href {https://doi.org/10.1103/PhysRevLett.125.017401} {\bibfield  {journal} {\bibinfo  {journal} {Phys. Rev. Lett.}\ }\textbf {\bibinfo {volume} {125}},\ \bibinfo {pages} {017401} (\bibinfo {year} {2020})}\BibitemShut {NoStop}%
\bibitem [{\citenamefont {Paoletta}\ and\ \citenamefont {Demkov}(2021)}]{Paoletta2021}%
  \BibitemOpen
  \bibfield  {author} {\bibinfo {author} {\bibfnamefont {T.}~\bibnamefont {Paoletta}}\ and\ \bibinfo {author} {\bibfnamefont {A.~A.}\ \bibnamefont {Demkov}},\ }\bibfield  {title} {\bibinfo {title} {Pockels effect in low-temperature rhombohedral $\mbox{BaTiO}_{3}$},\ }\href {https://doi.org/10.1103/PhysRevB.103.014303} {\bibfield  {journal} {\bibinfo  {journal} {Phys. Rev. B}\ }\textbf {\bibinfo {volume} {103}},\ \bibinfo {pages} {014303} (\bibinfo {year} {2021})}\BibitemShut {NoStop}%
\bibitem [{\citenamefont {Jiang}\ \emph {et~al.}(2024)\citenamefont {Jiang}, \citenamefont {Xiang}, \citenamefont {Bellaiche},\ and\ \citenamefont {Paillard}}]{jiang2024_2D}%
  \BibitemOpen
  \bibfield  {author} {\bibinfo {author} {\bibfnamefont {Z.}~\bibnamefont {Jiang}}, \bibinfo {author} {\bibfnamefont {H.}~\bibnamefont {Xiang}}, \bibinfo {author} {\bibfnamefont {L.}~\bibnamefont {Bellaiche}},\ and\ \bibinfo {author} {\bibfnamefont {C.}~\bibnamefont {Paillard}},\ }\bibfield  {title} {\bibinfo {title} {Electro-optic properties from ab initio calculations in two-dimensional materials},\ }\href {https://doi.org/10.1103/PhysRevB.109.165414} {\bibfield  {journal} {\bibinfo  {journal} {Phys. Rev. B}\ }\textbf {\bibinfo {volume} {109}},\ \bibinfo {pages} {165414} (\bibinfo {year} {2024})}\BibitemShut {NoStop}%
\bibitem [{\citenamefont {Zhang}\ \emph {et~al.}(2024)\citenamefont {Zhang}, \citenamefont {Di}, \citenamefont {Paillard}, \citenamefont {Bellaiche},\ and\ \citenamefont {Jiang}}]{Zhang2024}%
  \BibitemOpen
  \bibfield  {author} {\bibinfo {author} {\bibfnamefont {Z.}~\bibnamefont {Zhang}}, \bibinfo {author} {\bibfnamefont {X.}~\bibnamefont {Di}}, \bibinfo {author} {\bibfnamefont {C.}~\bibnamefont {Paillard}}, \bibinfo {author} {\bibfnamefont {L.}~\bibnamefont {Bellaiche}},\ and\ \bibinfo {author} {\bibfnamefont {Z.}~\bibnamefont {Jiang}},\ }\bibfield  {title} {\bibinfo {title} {Giant electro-optic and elasto-optic effects in ferroelectric $\mbox{NbOI}_{2}$},\ }\href {https://doi.org/10.1103/PhysRevB.110.L100101} {\bibfield  {journal} {\bibinfo  {journal} {Phys. Rev. B}\ }\textbf {\bibinfo {volume} {110}},\ \bibinfo {pages} {L100101} (\bibinfo {year} {2024})}\BibitemShut {NoStop}%
\bibitem [{\citenamefont {Zgonik}\ \emph {et~al.}(1994)\citenamefont {Zgonik}, \citenamefont {Bernasconi}, \citenamefont {Duelli}, \citenamefont {Schlesser}, \citenamefont {G{\"u}nter}, \citenamefont {Garrett}, \citenamefont {Rytz}, \citenamefont {Zhu},\ and\ \citenamefont {Wu}}]{zgonik1994}%
  \BibitemOpen
  \bibfield  {author} {\bibinfo {author} {\bibfnamefont {M.}~\bibnamefont {Zgonik}}, \bibinfo {author} {\bibfnamefont {P.}~\bibnamefont {Bernasconi}}, \bibinfo {author} {\bibfnamefont {M.}~\bibnamefont {Duelli}}, \bibinfo {author} {\bibfnamefont {R.}~\bibnamefont {Schlesser}}, \bibinfo {author} {\bibfnamefont {P.}~\bibnamefont {G{\"u}nter}}, \bibinfo {author} {\bibfnamefont {M.~H.}\ \bibnamefont {Garrett}}, \bibinfo {author} {\bibfnamefont {D.}~\bibnamefont {Rytz}}, \bibinfo {author} {\bibfnamefont {Y.}~\bibnamefont {Zhu}},\ and\ \bibinfo {author} {\bibfnamefont {X.}~\bibnamefont {Wu}},\ }\bibfield  {title} {\bibinfo {title} {Dielectric, elastic, piezoelectric, electro-optic, and elasto-optic tensors of $\mbox{BaTiO}_{3}$ crystals},\ }\href {https://doi.org/10.1103/PhysRevB.50.5941} {\bibfield  {journal} {\bibinfo  {journal} {Phys. Rev. B}\ }\textbf {\bibinfo {volume} {50}},\ \bibinfo {pages} {5941} (\bibinfo {year} {1994})}\BibitemShut {NoStop}%
\bibitem [{\citenamefont {Guthrie}\ and\ \citenamefont {Corbett}(1981)}]{guthrie1981}%
  \BibitemOpen
  \bibfield  {author} {\bibinfo {author} {\bibfnamefont {D.~H.}\ \bibnamefont {Guthrie}}\ and\ \bibinfo {author} {\bibfnamefont {J.~D.}\ \bibnamefont {Corbett}},\ }\bibfield  {title} {\bibinfo {title} {Synthesis and structure of an infinite-chain form of $\mbox{ZrI}_{2}$ ($\alpha$)},\ }\href {https://doi.org/10.1016/0022-4596(81)90092-X} {\bibfield  {journal} {\bibinfo  {journal} {J. Solid State Chem.}\ }\textbf {\bibinfo {volume} {37}},\ \bibinfo {pages} {256} (\bibinfo {year} {1981})}\BibitemShut {NoStop}%
\bibitem [{\citenamefont {Liu}\ \emph {et~al.}(2019)\citenamefont {Liu}, \citenamefont {Yang}, \citenamefont {Hu}, \citenamefont {Zhao}, \citenamefont {Chen},\ and\ \citenamefont {Ren}}]{Liu2019}%
  \BibitemOpen
  \bibfield  {author} {\bibinfo {author} {\bibfnamefont {X.}~\bibnamefont {Liu}}, \bibinfo {author} {\bibfnamefont {Y.}~\bibnamefont {Yang}}, \bibinfo {author} {\bibfnamefont {T.}~\bibnamefont {Hu}}, \bibinfo {author} {\bibfnamefont {G.}~\bibnamefont {Zhao}}, \bibinfo {author} {\bibfnamefont {C.}~\bibnamefont {Chen}},\ and\ \bibinfo {author} {\bibfnamefont {W.}~\bibnamefont {Ren}},\ }\bibfield  {title} {\bibinfo {title} {Vertical ferroelectric switching by in-plane sliding of two-dimensional bilayer $\mbox{WTe}_{2}$},\ }\href {https://doi.org/10.1039/C9NR05404A} {\bibfield  {journal} {\bibinfo  {journal} {Nanoscale}\ }\textbf {\bibinfo {volume} {11}},\ \bibinfo {pages} {18575} (\bibinfo {year} {2019})}\BibitemShut {NoStop}%
\bibitem [{\citenamefont {Tang}\ \emph {et~al.}(2025)\citenamefont {Tang}, \citenamefont {Tian}, \citenamefont {Ouyang}, \citenamefont {Pan},\ and\ \citenamefont {Chen}}]{Tang2025}%
  \BibitemOpen
  \bibfield  {author} {\bibinfo {author} {\bibfnamefont {C.}~\bibnamefont {Tang}}, \bibinfo {author} {\bibfnamefont {Z.}~\bibnamefont {Tian}}, \bibinfo {author} {\bibfnamefont {T.}~\bibnamefont {Ouyang}}, \bibinfo {author} {\bibfnamefont {A.}~\bibnamefont {Pan}},\ and\ \bibinfo {author} {\bibfnamefont {M.}~\bibnamefont {Chen}},\ }\bibfield  {title} {\bibinfo {title} {Combining intrinsic and sliding-induced polarizations for multistates in two-dimensional ferroelectrics},\ }\href {https://doi.org/10.1103/PhysRevB.111.L081407} {\bibfield  {journal} {\bibinfo  {journal} {Phys. Rev. B}\ }\textbf {\bibinfo {volume} {111}},\ \bibinfo {pages} {L081407} (\bibinfo {year} {2025})}\BibitemShut {NoStop}%
\bibitem [{\citenamefont {Nye}(1985)}]{Nye1985}%
  \BibitemOpen
  \bibfield  {author} {\bibinfo {author} {\bibfnamefont {J.~F.}\ \bibnamefont {Nye}},\ }\href@noop {} {\emph {\bibinfo {title} {Physical Properties of Crystals: Their Representation by Tensors and Matrices}}}\ (\bibinfo  {publisher} {Oxford University Press, Oxford},\ \bibinfo {year} {1985})\BibitemShut {NoStop}%
\bibitem [{\citenamefont {Weber}(2002)}]{Weber2002}%
  \BibitemOpen
  \bibfield  {author} {\bibinfo {author} {\bibfnamefont {M.~J.}\ \bibnamefont {Weber}},\ }\href@noop {} {\emph {\bibinfo {title} {Handbook of optical materials}}}\ (\bibinfo  {publisher} {CRC Press, Boca Raton, FL},\ \bibinfo {year} {2002})\BibitemShut {NoStop}%
\bibitem [{\citenamefont {Jiang}\ \emph {et~al.}(2022)\citenamefont {Jiang}, \citenamefont {Paillard}, \citenamefont {Churchill}, \citenamefont {Xia}, \citenamefont {Zhang}, \citenamefont {Xiang},\ and\ \citenamefont {Bellaiche}}]{Jiang2022}%
  \BibitemOpen
  \bibfield  {author} {\bibinfo {author} {\bibfnamefont {Z.}~\bibnamefont {Jiang}}, \bibinfo {author} {\bibfnamefont {C.}~\bibnamefont {Paillard}}, \bibinfo {author} {\bibfnamefont {H.~O.~H.}\ \bibnamefont {Churchill}}, \bibinfo {author} {\bibfnamefont {M.}~\bibnamefont {Xia}}, \bibinfo {author} {\bibfnamefont {S.}~\bibnamefont {Zhang}}, \bibinfo {author} {\bibfnamefont {H.}~\bibnamefont {Xiang}},\ and\ \bibinfo {author} {\bibfnamefont {L.}~\bibnamefont {Bellaiche}},\ }\bibfield  {title} {\bibinfo {title} {Large linear and nonlinear electro-optic coefficients in two-dimensional ferroelectrics},\ }\href {https://doi.org/10.1103/PhysRevB.106.L081404} {\bibfield  {journal} {\bibinfo  {journal} {Phys. Rev. B}\ }\textbf {\bibinfo {volume} {106}},\ \bibinfo {pages} {L081404} (\bibinfo {year} {2022})}\BibitemShut {NoStop}%
\bibitem [{\citenamefont {Delodovici}\ \emph {et~al.}(2023)\citenamefont {Delodovici}, \citenamefont {Atkinson}, \citenamefont {Xu}, \citenamefont {Janolin}, \citenamefont {Alpay},\ and\ \citenamefont {Paillard}}]{delodovici2023}%
  \BibitemOpen
  \bibfield  {author} {\bibinfo {author} {\bibfnamefont {F.}~\bibnamefont {Delodovici}}, \bibinfo {author} {\bibfnamefont {C.}~\bibnamefont {Atkinson}}, \bibinfo {author} {\bibfnamefont {R.}~\bibnamefont {Xu}}, \bibinfo {author} {\bibfnamefont {P.-E.}\ \bibnamefont {Janolin}}, \bibinfo {author} {\bibfnamefont {S.~P.}\ \bibnamefont {Alpay}},\ and\ \bibinfo {author} {\bibfnamefont {C.}~\bibnamefont {Paillard}},\ }\bibfield  {title} {\bibinfo {title} {Engineering the electro-optic effect in $\mbox{HfO}_{2}$ and $\mbox{ZrO}_{2}$ through strain and polarization control},\ }\href {https://doi.org/10.1063/5.0158909} {\bibfield  {journal} {\bibinfo  {journal} {J. Appl. Phys.}\ }\textbf {\bibinfo {volume} {134}},\ \bibinfo {pages} {055108} (\bibinfo {year} {2023})}\BibitemShut {NoStop}%
\bibitem [{\citenamefont {Ding}\ \emph {et~al.}(2024)\citenamefont {Ding}, \citenamefont {Huang}, \citenamefont {Wu},\ and\ \citenamefont {Zhou}}]{Ding2024}%
  \BibitemOpen
  \bibfield  {author} {\bibinfo {author} {\bibfnamefont {Y.-m.}\ \bibnamefont {Ding}}, \bibinfo {author} {\bibfnamefont {A.}~\bibnamefont {Huang}}, \bibinfo {author} {\bibfnamefont {Y.}~\bibnamefont {Wu}},\ and\ \bibinfo {author} {\bibfnamefont {L.}~\bibnamefont {Zhou}},\ }\bibfield  {title} {\bibinfo {title} {Strain-induced ferroelectric phase transition and second-harmonic generation enhancement in $\mbox{NbOCl}_{2}$ monolayer},\ }\href {https://doi.org/10.1063/5.0235837} {\bibfield  {journal} {\bibinfo  {journal} {Appl. Phys. Lett.}\ }\textbf {\bibinfo {volume} {125}},\ \bibinfo {pages} {151902} (\bibinfo {year} {2024})}\BibitemShut {NoStop}%
\bibitem [{\citenamefont {Saleh}\ and\ \citenamefont {Teich}(2007)}]{saleh2019}%
  \BibitemOpen
  \bibfield  {author} {\bibinfo {author} {\bibfnamefont {B.~E.~A.}\ \bibnamefont {Saleh}}\ and\ \bibinfo {author} {\bibfnamefont {M.~C.}\ \bibnamefont {Teich}},\ }\href@noop {} {\emph {\bibinfo {title} {Fundamentals of Photonics}}}\ (\bibinfo  {publisher} {Wiley, Hoboken, NJ},\ \bibinfo {year} {2007})\BibitemShut {NoStop}%
\bibitem [{\citenamefont {Eichenfield}\ \emph {et~al.}(2009)\citenamefont {Eichenfield}, \citenamefont {Chan}, \citenamefont {Camacho}, \citenamefont {Vahala},\ and\ \citenamefont {Painter}}]{eichenfield2009}%
  \BibitemOpen
  \bibfield  {author} {\bibinfo {author} {\bibfnamefont {M.}~\bibnamefont {Eichenfield}}, \bibinfo {author} {\bibfnamefont {J.}~\bibnamefont {Chan}}, \bibinfo {author} {\bibfnamefont {R.~M.}\ \bibnamefont {Camacho}}, \bibinfo {author} {\bibfnamefont {K.~J.}\ \bibnamefont {Vahala}},\ and\ \bibinfo {author} {\bibfnamefont {O.}~\bibnamefont {Painter}},\ }\bibfield  {title} {\bibinfo {title} {Optomechanical crystals},\ }\href {https://doi.org/10.1038/nature08524} {\bibfield  {journal} {\bibinfo  {journal} {Nature (London)}\ }\textbf {\bibinfo {volume} {462}},\ \bibinfo {pages} {78} (\bibinfo {year} {2009})}\BibitemShut {NoStop}%
\bibitem [{\citenamefont {Yasuda}\ \emph {et~al.}(2024)\citenamefont {Yasuda}, \citenamefont {Zalys-Geller}, \citenamefont {Wang}, \citenamefont {Bennett}, \citenamefont {Cheema}, \citenamefont {Watanabe}, \citenamefont {Taniguchi}, \citenamefont {Kaxiras}, \citenamefont {Jarillo-Herrero},\ and\ \citenamefont {Ashoori}}]{Yasuda2024}%
  \BibitemOpen
  \bibfield  {author} {\bibinfo {author} {\bibfnamefont {K.}~\bibnamefont {Yasuda}}, \bibinfo {author} {\bibfnamefont {E.}~\bibnamefont {Zalys-Geller}}, \bibinfo {author} {\bibfnamefont {X.}~\bibnamefont {Wang}}, \bibinfo {author} {\bibfnamefont {D.}~\bibnamefont {Bennett}}, \bibinfo {author} {\bibfnamefont {S.~S.}\ \bibnamefont {Cheema}}, \bibinfo {author} {\bibfnamefont {K.}~\bibnamefont {Watanabe}}, \bibinfo {author} {\bibfnamefont {T.}~\bibnamefont {Taniguchi}}, \bibinfo {author} {\bibfnamefont {E.}~\bibnamefont {Kaxiras}}, \bibinfo {author} {\bibfnamefont {P.}~\bibnamefont {Jarillo-Herrero}},\ and\ \bibinfo {author} {\bibfnamefont {R.}~\bibnamefont {Ashoori}},\ }\bibfield  {title} {\bibinfo {title} {Ultrafast high-endurance memory based on sliding ferroelectrics},\ }\href {https://doi.org/10.1126/science.adp3575} {\bibfield  {journal} {\bibinfo  {journal} {Science}\ }\textbf {\bibinfo {volume} {385}},\ \bibinfo {pages} {53} (\bibinfo {year} {2024})}\BibitemShut {NoStop}%
\bibitem [{\citenamefont {Bian}\ \emph {et~al.}(2024)\citenamefont {Bian}, \citenamefont {He}, \citenamefont {Pan}, \citenamefont {Li}, \citenamefont {Cao}, \citenamefont {Meng}, \citenamefont {Chen}, \citenamefont {Liu}, \citenamefont {Zhong}, \citenamefont {Li},\ and\ \citenamefont {Liu}}]{Bian2024}%
  \BibitemOpen
  \bibfield  {author} {\bibinfo {author} {\bibfnamefont {R.}~\bibnamefont {Bian}}, \bibinfo {author} {\bibfnamefont {R.}~\bibnamefont {He}}, \bibinfo {author} {\bibfnamefont {E.}~\bibnamefont {Pan}}, \bibinfo {author} {\bibfnamefont {Z.}~\bibnamefont {Li}}, \bibinfo {author} {\bibfnamefont {G.}~\bibnamefont {Cao}}, \bibinfo {author} {\bibfnamefont {P.}~\bibnamefont {Meng}}, \bibinfo {author} {\bibfnamefont {J.}~\bibnamefont {Chen}}, \bibinfo {author} {\bibfnamefont {Q.}~\bibnamefont {Liu}}, \bibinfo {author} {\bibfnamefont {Z.}~\bibnamefont {Zhong}}, \bibinfo {author} {\bibfnamefont {W.}~\bibnamefont {Li}},\ and\ \bibinfo {author} {\bibfnamefont {F.}~\bibnamefont {Liu}},\ }\bibfield  {title} {\bibinfo {title} {Developing fatigue-resistant ferroelectrics using interlayer sliding switching},\ }\href {https://doi.org/10.1126/science.ado1744} {\bibfield  {journal} {\bibinfo  {journal} {Science}\ }\textbf {\bibinfo {volume} {385}},\ \bibinfo {pages} {57} (\bibinfo {year} {2024})}\BibitemShut {NoStop}%
\bibitem [{\citenamefont {Togo}\ \emph {et~al.}(2023)\citenamefont {Togo}, \citenamefont {Chaput}, \citenamefont {Tadano},\ and\ \citenamefont {Tanaka}}]{Togo2023}%
  \BibitemOpen
  \bibfield  {author} {\bibinfo {author} {\bibfnamefont {A.}~\bibnamefont {Togo}}, \bibinfo {author} {\bibfnamefont {L.}~\bibnamefont {Chaput}}, \bibinfo {author} {\bibfnamefont {T.}~\bibnamefont {Tadano}},\ and\ \bibinfo {author} {\bibfnamefont {I.}~\bibnamefont {Tanaka}},\ }\bibfield  {title} {\bibinfo {title} {Implementation strategies in phonopy and phono3py},\ }\href {https://doi.org/https://doi.org/10.1088/1361-648X/acd831} {\bibfield  {journal} {\bibinfo  {journal} {J. Phys.: Condens. Matter}\ }\textbf {\bibinfo {volume} {35}},\ \bibinfo {pages} {353001} (\bibinfo {year} {2023})}\BibitemShut {NoStop}%
\bibitem [{\citenamefont {Togo}(2023)}]{Togo2023-JPSJ}%
  \BibitemOpen
  \bibfield  {author} {\bibinfo {author} {\bibfnamefont {A.}~\bibnamefont {Togo}},\ }\bibfield  {title} {\bibinfo {title} {First-principles phonon calculations with phonopy and phono3py},\ }\href {https://doi.org/https://doi.org/10.7566/JPSJ.92.012001} {\bibfield  {journal} {\bibinfo  {journal} {J. Phys. Soc. Jpn.}\ }\textbf {\bibinfo {volume} {92}},\ \bibinfo {pages} {012001} (\bibinfo {year} {2023})}\BibitemShut {NoStop}%
\bibitem [{\citenamefont {Kresse}\ and\ \citenamefont {Furthm{\"u}ller}(1996)}]{Kresse1996}%
  \BibitemOpen
  \bibfield  {author} {\bibinfo {author} {\bibfnamefont {G.}~\bibnamefont {Kresse}}\ and\ \bibinfo {author} {\bibfnamefont {J.}~\bibnamefont {Furthm{\"u}ller}},\ }\bibfield  {title} {\bibinfo {title} {Efficiency of ab-initio total energy calculations for metals and semiconductors using a plane-wave basis set},\ }\href {https://doi.org/10.1016/0927-0256(96)00008-0} {\bibfield  {journal} {\bibinfo  {journal} {Comput. Mater. Sci.}\ }\textbf {\bibinfo {volume} {6}},\ \bibinfo {pages} {15} (\bibinfo {year} {1996})}\BibitemShut {NoStop}%
\bibitem [{\citenamefont {Bl{\"o}chl}(1994)}]{Blochl1994}%
  \BibitemOpen
  \bibfield  {author} {\bibinfo {author} {\bibfnamefont {P.~E.}\ \bibnamefont {Bl{\"o}chl}},\ }\bibfield  {title} {\bibinfo {title} {Projector augmented-wave method},\ }\href {https://doi.org/10.1103/PhysRevB.50.17953} {\bibfield  {journal} {\bibinfo  {journal} {Phys. Rev. B}\ }\textbf {\bibinfo {volume} {50}},\ \bibinfo {pages} {17953} (\bibinfo {year} {1994})}\BibitemShut {NoStop}%
\bibitem [{\citenamefont {Perdew}\ \emph {et~al.}(1996)\citenamefont {Perdew}, \citenamefont {Burke},\ and\ \citenamefont {Ernzerhof}}]{Perdew1996}%
  \BibitemOpen
  \bibfield  {author} {\bibinfo {author} {\bibfnamefont {J.~P.}\ \bibnamefont {Perdew}}, \bibinfo {author} {\bibfnamefont {K.}~\bibnamefont {Burke}},\ and\ \bibinfo {author} {\bibfnamefont {M.}~\bibnamefont {Ernzerhof}},\ }\bibfield  {title} {\bibinfo {title} {Generalized gradient approximation made simple},\ }\href {https://doi.org/10.1103/PhysRevLett.77.3865} {\bibfield  {journal} {\bibinfo  {journal} {Phys. Rev. Lett.}\ }\textbf {\bibinfo {volume} {77}},\ \bibinfo {pages} {3865} (\bibinfo {year} {1996})}\BibitemShut {NoStop}%
\bibitem [{\citenamefont {Grimme}\ \emph {et~al.}(2011)\citenamefont {Grimme}, \citenamefont {Ehrlich},\ and\ \citenamefont {Goerigk}}]{Grimme2011}%
  \BibitemOpen
  \bibfield  {author} {\bibinfo {author} {\bibfnamefont {S.}~\bibnamefont {Grimme}}, \bibinfo {author} {\bibfnamefont {S.}~\bibnamefont {Ehrlich}},\ and\ \bibinfo {author} {\bibfnamefont {L.}~\bibnamefont {Goerigk}},\ }\bibfield  {title} {\bibinfo {title} {Effect of the damping function in dispersion corrected density functional theory},\ }\href {https://doi.org/10.1002/jcc.21759} {\bibfield  {journal} {\bibinfo  {journal} {J. Comput. Chem.}\ }\textbf {\bibinfo {volume} {32}},\ \bibinfo {pages} {1456} (\bibinfo {year} {2011})}\BibitemShut {NoStop}%
\bibitem [{\citenamefont {Grimme}\ \emph {et~al.}(2010)\citenamefont {Grimme}, \citenamefont {Antony}, \citenamefont {Ehrlich},\ and\ \citenamefont {Krieg}}]{Grimme2010}%
  \BibitemOpen
  \bibfield  {author} {\bibinfo {author} {\bibfnamefont {S.}~\bibnamefont {Grimme}}, \bibinfo {author} {\bibfnamefont {J.}~\bibnamefont {Antony}}, \bibinfo {author} {\bibfnamefont {S.}~\bibnamefont {Ehrlich}},\ and\ \bibinfo {author} {\bibfnamefont {H.}~\bibnamefont {Krieg}},\ }\bibfield  {title} {\bibinfo {title} {A consistent and accurate ab initio parametrization of density functional dispersion correction ($\mbox{DFT-D}$) for the 94 elements $\mbox{H-Pu}$},\ }\href {https://doi.org/10.1063/1.3382344} {\bibfield  {journal} {\bibinfo  {journal} {J. Chem. Phys.}\ }\textbf {\bibinfo {volume} {132}},\ \bibinfo {pages} {154104} (\bibinfo {year} {2010})}\BibitemShut {NoStop}%
\bibitem [{\citenamefont {King-Smith}\ and\ \citenamefont {Vanderbilt}(1993)}]{King-Smith1993}%
  \BibitemOpen
  \bibfield  {author} {\bibinfo {author} {\bibfnamefont {R.~D.}\ \bibnamefont {King-Smith}}\ and\ \bibinfo {author} {\bibfnamefont {D.}~\bibnamefont {Vanderbilt}},\ }\bibfield  {title} {\bibinfo {title} {Theory of polarization of crystalline solids},\ }\href {https://doi.org/10.1103/PhysRevB.47.1651} {\bibfield  {journal} {\bibinfo  {journal} {Phys. Rev. B}\ }\textbf {\bibinfo {volume} {47}},\ \bibinfo {pages} {1651} (\bibinfo {year} {1993})}\BibitemShut {NoStop}%
\bibitem [{\citenamefont {Resta}(1994)}]{Resta1994}%
  \BibitemOpen
  \bibfield  {author} {\bibinfo {author} {\bibfnamefont {R.}~\bibnamefont {Resta}},\ }\bibfield  {title} {\bibinfo {title} {Macroscopic polarization in crystalline dielectrics: the geometric phase approach},\ }\href {https://doi.org/10.1103/RevModPhys.66.899} {\bibfield  {journal} {\bibinfo  {journal} {Rev. Mod. Phys.}\ }\textbf {\bibinfo {volume} {66}},\ \bibinfo {pages} {899} (\bibinfo {year} {1994})}\BibitemShut {NoStop}%
\bibitem [{\citenamefont {Henkelman}\ \emph {et~al.}(2000)\citenamefont {Henkelman}, \citenamefont {Uberuaga},\ and\ \citenamefont {J{\'o}nsson}}]{Henkelman2000}%
  \BibitemOpen
  \bibfield  {author} {\bibinfo {author} {\bibfnamefont {G.}~\bibnamefont {Henkelman}}, \bibinfo {author} {\bibfnamefont {B.~P.}\ \bibnamefont {Uberuaga}},\ and\ \bibinfo {author} {\bibfnamefont {H.}~\bibnamefont {J{\'o}nsson}},\ }\bibfield  {title} {\bibinfo {title} {A climbing image nudged elastic band method for finding saddle points and minimum energy paths},\ }\href {https://doi.org/10.1063/1.1329672} {\bibfield  {journal} {\bibinfo  {journal} {J. Chem. Phys.}\ }\textbf {\bibinfo {volume} {113}},\ \bibinfo {pages} {9901} (\bibinfo {year} {2000})}\BibitemShut {NoStop}%
\bibitem [{\citenamefont {Henkelman}\ and\ \citenamefont {J{\'o}nsson}(2000)}]{Henkelman2020-2}%
  \BibitemOpen
  \bibfield  {author} {\bibinfo {author} {\bibfnamefont {G.}~\bibnamefont {Henkelman}}\ and\ \bibinfo {author} {\bibfnamefont {H.}~\bibnamefont {J{\'o}nsson}},\ }\bibfield  {title} {\bibinfo {title} {Improved tangent estimate in the nudged elastic band method for finding minimum energy paths and saddle points},\ }\href {https://doi.org/10.1063/1.1323224} {\bibfield  {journal} {\bibinfo  {journal} {J. Chem. Phys.}\ }\textbf {\bibinfo {volume} {113}},\ \bibinfo {pages} {9978} (\bibinfo {year} {2000})}\BibitemShut {NoStop}%
\bibitem [{\citenamefont {Gonze}\ \emph {et~al.}(2002)\citenamefont {Gonze}, \citenamefont {Beuken}, \citenamefont {Caracas}, \citenamefont {Detraux}, \citenamefont {Fuchs}, \citenamefont {Rignanese}, \citenamefont {Sindic}, \citenamefont {Verstraete}, \citenamefont {Zerah}, \citenamefont {Jollet}, \citenamefont {Torrent}, \citenamefont {Roy}, \citenamefont {Mikami}, \citenamefont {Ghosez}, \citenamefont {Raty},\ and\ \citenamefont {Allan}}]{Gonze2002}%
  \BibitemOpen
  \bibfield  {author} {\bibinfo {author} {\bibfnamefont {X.}~\bibnamefont {Gonze}}, \bibinfo {author} {\bibfnamefont {J.-M.}\ \bibnamefont {Beuken}}, \bibinfo {author} {\bibfnamefont {R.}~\bibnamefont {Caracas}}, \bibinfo {author} {\bibfnamefont {F.}~\bibnamefont {Detraux}}, \bibinfo {author} {\bibfnamefont {M.}~\bibnamefont {Fuchs}}, \bibinfo {author} {\bibfnamefont {G.-M.}\ \bibnamefont {Rignanese}}, \bibinfo {author} {\bibfnamefont {L.}~\bibnamefont {Sindic}}, \bibinfo {author} {\bibfnamefont {M.}~\bibnamefont {Verstraete}}, \bibinfo {author} {\bibfnamefont {G.}~\bibnamefont {Zerah}}, \bibinfo {author} {\bibfnamefont {F.}~\bibnamefont {Jollet}}, \bibinfo {author} {\bibfnamefont {M.}~\bibnamefont {Torrent}}, \bibinfo {author} {\bibfnamefont {A.}~\bibnamefont {Roy}}, \bibinfo {author} {\bibfnamefont {M.}~\bibnamefont {Mikami}}, \bibinfo {author} {\bibfnamefont {P.}~\bibnamefont {Ghosez}}, \bibinfo {author} {\bibfnamefont {J.-Y.}\ \bibnamefont {Raty}},\ and\ \bibinfo {author} {\bibfnamefont {D.~C.}\
  \bibnamefont {Allan}},\ }\bibfield  {title} {\bibinfo {title} {First-principles computation of material properties: the $\mbox{ABINIT}$ software project},\ }\href {https://doi.org/10.1016/S0927-0256(02)00325-7} {\bibfield  {journal} {\bibinfo  {journal} {Comput. Mater. Sci.}\ }\textbf {\bibinfo {volume} {25}},\ \bibinfo {pages} {478} (\bibinfo {year} {2002})}\BibitemShut {NoStop}%
\bibitem [{\citenamefont {Hamann}(2013)}]{hamann2013}%
  \BibitemOpen
  \bibfield  {author} {\bibinfo {author} {\bibfnamefont {D.~R.}\ \bibnamefont {Hamann}},\ }\bibfield  {title} {\bibinfo {title} {Optimized norm-conserving vanderbilt pseudopotentials},\ }\href {https://doi.org/10.1103/PhysRevB.88.085117} {\bibfield  {journal} {\bibinfo  {journal} {Phys. Rev. B}\ }\textbf {\bibinfo {volume} {88}},\ \bibinfo {pages} {085117} (\bibinfo {year} {2013})}\BibitemShut {NoStop}%
\bibitem [{\citenamefont {Laturia}\ \emph {et~al.}(2018)\citenamefont {Laturia}, \citenamefont {Van~de Put},\ and\ \citenamefont {Vandenberghe}}]{laturia2018}%
  \BibitemOpen
  \bibfield  {author} {\bibinfo {author} {\bibfnamefont {A.}~\bibnamefont {Laturia}}, \bibinfo {author} {\bibfnamefont {M.~L.}\ \bibnamefont {Van~de Put}},\ and\ \bibinfo {author} {\bibfnamefont {W.~G.}\ \bibnamefont {Vandenberghe}},\ }\bibfield  {title} {\bibinfo {title} {Dielectric properties of hexagonal boron nitride and transition metal dichalcogenides: from monolayer to bulk},\ }\href {https://doi.org/10.1038/s41699-018-0050-x} {\bibfield  {journal} {\bibinfo  {journal} {npj 2D Mater. Appl.}\ }\textbf {\bibinfo {volume} {2}},\ \bibinfo {pages} {6} (\bibinfo {year} {2018})}\BibitemShut {NoStop}%
\bibitem [{\citenamefont {Bennett}\ \emph {et~al.}(2023)\citenamefont {Bennett}, \citenamefont {Chaudhary}, \citenamefont {Slager}, \citenamefont {Bousquet},\ and\ \citenamefont {Ghosez}}]{bennett2023}%
  \BibitemOpen
  \bibfield  {author} {\bibinfo {author} {\bibfnamefont {D.}~\bibnamefont {Bennett}}, \bibinfo {author} {\bibfnamefont {G.}~\bibnamefont {Chaudhary}}, \bibinfo {author} {\bibfnamefont {R.-J.}\ \bibnamefont {Slager}}, \bibinfo {author} {\bibfnamefont {E.}~\bibnamefont {Bousquet}},\ and\ \bibinfo {author} {\bibfnamefont {P.}~\bibnamefont {Ghosez}},\ }\bibfield  {title} {\bibinfo {title} {Polar meron-antimeron networks in strained and twisted bilayers},\ }\href {https://doi.org/10.1038/s41467-023-37337-8} {\bibfield  {journal} {\bibinfo  {journal} {Nat. Commun.}\ }\textbf {\bibinfo {volume} {14}},\ \bibinfo {pages} {1629} (\bibinfo {year} {2023})}\BibitemShut {NoStop}%
\bibitem [{\citenamefont {Wang}\ \emph {et~al.}(2023)\citenamefont {Wang}, \citenamefont {Gui},\ and\ \citenamefont {Huang}}]{Wang2023}%
  \BibitemOpen
  \bibfield  {author} {\bibinfo {author} {\bibfnamefont {Z.}~\bibnamefont {Wang}}, \bibinfo {author} {\bibfnamefont {Z.}~\bibnamefont {Gui}},\ and\ \bibinfo {author} {\bibfnamefont {L.}~\bibnamefont {Huang}},\ }\bibfield  {title} {\bibinfo {title} {Sliding ferroelectricity in bilayer honeycomb structures: A first-principles study},\ }\href {https://doi.org/10.1103/PhysRevB.107.035426} {\bibfield  {journal} {\bibinfo  {journal} {Phys. Rev. B}\ }\textbf {\bibinfo {volume} {107}},\ \bibinfo {pages} {035426} (\bibinfo {year} {2023})}\BibitemShut {NoStop}%
\bibitem [{\citenamefont {Behzad}(2024)}]{behzad2024monolayer}%
  \BibitemOpen
  \bibfield  {author} {\bibinfo {author} {\bibfnamefont {S.}~\bibnamefont {Behzad}},\ }\bibfield  {title} {\bibinfo {title} {Monolayer and bilayer $\mbox{BP}$ as efficient optoelectronic materials in visible and ultraviolet regions},\ }\href {https://doi.org/10.1016/j.rinp.2024.108048} {\bibfield  {journal} {\bibinfo  {journal} {Results Phys.}\ }\textbf {\bibinfo {volume} {67}},\ \bibinfo {pages} {108048} (\bibinfo {year} {2024})}\BibitemShut {NoStop}%
\bibitem [{\citenamefont {Zhong}\ \emph {et~al.}(2011)\citenamefont {Zhong}, \citenamefont {Yap}, \citenamefont {Pandey},\ and\ \citenamefont {Karna}}]{Zhong2011}%
  \BibitemOpen
  \bibfield  {author} {\bibinfo {author} {\bibfnamefont {X.}~\bibnamefont {Zhong}}, \bibinfo {author} {\bibfnamefont {Y.~K.}\ \bibnamefont {Yap}}, \bibinfo {author} {\bibfnamefont {R.}~\bibnamefont {Pandey}},\ and\ \bibinfo {author} {\bibfnamefont {S.~P.}\ \bibnamefont {Karna}},\ }\bibfield  {title} {\bibinfo {title} {First-principles study of strain-induced modulation of energy gaps of graphene/$\mbox{BN}$ and $\mbox{BN}$ bilayers},\ }\href {https://doi.org/10.1103/PhysRevB.83.193403} {\bibfield  {journal} {\bibinfo  {journal} {Phys. Rev. B}\ }\textbf {\bibinfo {volume} {83}},\ \bibinfo {pages} {193403} (\bibinfo {year} {2011})}\BibitemShut {NoStop}%
\end{thebibliography}%

\end{document}